\documentclass[12pt]{article}

\usepackage[utf8]{inputenc}
\usepackage{graphicx}
\usepackage{amsmath}
\usepackage{amssymb}
\usepackage{amsfonts}
\usepackage[inline]{enumitem}	
\usepackage{tensor}
\usepackage{braket}
\usepackage{lmodern}
\usepackage{setspace}
\usepackage[margin=2.5cm]{geometry}
\usepackage{amsthm}
\usepackage{mathtools}
\usepackage{color}
\usepackage{hyperref}
\usepackage{dsfont}
\usepackage{authblk}
\usepackage{cite}

\DeclareMathOperator{\tr}{tr}
\DeclareMathOperator{\sinc}{sinc}

\newcommand{\updated}[1]{{#1}}

\begin{document}

\title{Impact of relativity on particle localizability\\ and ground state entanglement}

\author[1,2,3]{Maria Papageorgiou}
\author[1,2,3]{Jason Pye}

\affil[1]{\normalsize Department of Applied Mathematics, University of Waterloo, Waterloo, Ontario, N2L 3G1, Canada}
\affil[2]{\normalsize Institute for Quantum Computing, University of Waterloo, Waterloo, Ontario, N2L 3G1, Canada}
\affil[3]{\normalsize Perimeter Institute for Theoretical Physics, 31 Caroline St. N., Waterloo, Ontario, N2L 2Y5, Canada}

\date{\normalsize \today}

\maketitle

\begin{abstract}
  Can a relativistic quantum field theory be consistently described as a theory of localizable particles? There are many known issues with such a description, indicating an answer in the negative. In this paper, we examine these obstructions by tracing how they (partially) subside in passing to an approximation of ordinary quantum mechanics in the non-relativistic regime. We undertake a recovery of the characteristic features of non-relativistic quantum mechanics beyond simply the Schr\"odinger equation. We find that once this is achieved, there are persisting issues in the localizability of particle states. A major focus is on the lingering discrepancy between two different localization schemes in quantum field theory. The non-relativistic approximation of the quantum field theory is achieved by introducing an ultraviolet cutoff set by the Compton scale. The other main undertaking of this paper is to quantify the fate of ground state entanglement and the Unruh effect in the non-relativistic regime. Observing that the Unruh temperature vanishes in the naive limit as the speed of light is taken to infinity motivates the question: is the Unruh effect relativistic? It happens that this is closely related to the former issues, as ground state entanglement causes obstructions to localizability through the Reeh-Schlieder theorem. 
\end{abstract}

\newpage

\tableofcontents

\newpage

\section{Introduction}

Much of the structure of quantum field theory (QFT) is predicated on the principle of locality.
Adherence to locality is pursuant to convictions rooted in relativity, and is achieved in QFT by the association of regions of spacetime with algebras of observables.
Although, by construction, the observables of QFT are local objects, one may also consider characterizing the spatial or spacetime features of a \emph{state}.
For example, if we have a single-particle state, how can we say that the particle is localized in a certain region of space?
It turns out that often such a characterization is obstructed by one of a collection of no-go theorems, which, for example, imply the absence of any suitable definition of a position operator or local number operators.
These difficulties seem to suggest that relativistic QFT cannot support an ontology in terms of localizable particles. The tension between the local nature of observables and the non-localizability of particle states has been studied in \cite{Colosi-Rovelli2008}.
Our aim here is to investigate relativity as the source of this tension, by examining the non-relativistic regime of a relativistic QFT.

What precisely are the difficulties with localized particle states in QFT?
The problem can be seen as a result of competing requirements that one would like to attribute to such states.
Perhaps the most basic requirement is that particles, despite not being able to carry labels, should be aggregable entities that can be counted, i.e., there should be an observable number operator acting on a corresponding Fock space.
Other important stipulations include that particles should persist in time (at least in free theories) and that the particle excitations (quanta) should exhibit the appropriate relativistic dispersion relation between mass, momentum, and energy \cite{Teller1997, Fraser2008}.
As Fraser writes:
\begin{quote}
Without the mass-energy relation, there would be no grounds for interpreting the eigenstates of total number operator $N$ as representing definite numbers of particles rather than merely more examples of discrete energy level states that are the hallmark of non-relativistic quantum mechanics. However, what special relativity gives, special relativity also takes away. For free systems, relativistic assumptions are required to obtain the result that quanta are not localizable in a finite region of space. \cite{FraserDraft}
\end{quote}

An instinctive means to characterize the localizability of a state is to concoct a position operator in QFT analogous to that of non-relativistic quantum mechanics (NRQM).
The spectral projections of such an operator can be used to determine the probability of finding the particle in a certain region of space, given by $|\psi(x)|^2$ for a state $\ket{\psi}$.
\updated{Unlike the momentum and energy operators, the Poincar\'e group does not naturally provide us with an operator that corresponds to a position observable.
Newton and Wigner \cite{newton1949localized} devised a position operator for elementary ``particles'' (i.e., irreducible representations of the Poincar\'e group).
However, it is well-known that this operator suffers from issues with superluminal signalling \cite{Fulling1989}.}
In a more general context, Malament's theorem \cite{Malament1996} demonstrates that the only projections satisfying a short list of reasonable requirements are trivial, suggesting that there cannot be any analogue of a position operator in a relativistic quantum theory.

Another possibility may be to construct a local number operator, in an attempt to count the number of particles in a particular region of space.
This is obstructed by a corollary to the Reeh-Schlieder theorem \cite{Reeh-Schlieder1961}, which has been argued is a result of entanglement in the vacuum state \cite{Redhead1995}. 
A more operational approach to describe the localizability of a state in QFT is in terms of expectation values of observables restricted to particular regions of spacetime.
A theorem of Hegerfeldt \cite{Hegerfeldt1998a,Hegerfeldt1998b,Hegerfeldt1974,Halvorson2001} demonstrates that fixed particle number states localized in some region will have an effect on expectation values of observables in spacelike separated regions.
Here we will investigate whether the incompatibility of these requirements is placated in the non-relativistic regime of the theory, where one expects to recover localizable particle states.

Why should one insist in characterizing the localizability of fixed particle-number states?
There are both phenomenological and ontological aspects to this question.
Of course, ultimately it is well-known that the notion of particle is observer-dependent.
Despite this, there are situations where one may be inclined to describe localizable particle states for a fixed observer.
For instance, high energy `particle physics' experiments (e.g., in particle colliders) as well as low energy quantum information experiments (e.g., single photon detectors) demonstrate particle-like phenomenology that should be accounted for, perhaps using suitable detector models.
Nevertheless, a successful detector model capturing this particle-like phenomenology may evade---but cannot answer---the ontological aspects of the question.
Philosophers of QFT have offered proposals about why we are allowed `particle talk without particle ontology' \cite{Halvorson-Clifton2002}.
If relativistic QFT does not admit a particle ontology, it is fair to ask: what do particle detectors detect?
Looking towards low energies, one finds the widespread applicability of NRQM, a theory in which particle states are localizable by means of their wavefunction.
This seems to imply that NRQM can support a particle ontology, so it is natural to ask whether one can make contact between the NRQM description of particles and some appropriate notion in the latent QFT.

Admittedly QFT amd NRQM are very different theories, both at the dynamical and kinematical level, and recovering features of one from the other cannot come with no cost.
The undertaking of this paper will be to illuminate this connection, by starting with a relativistic QFT and making suitable approximations to recover features of NRQM.
Departing from a relativistic QFT, we wish to clarify the measures that need to be taken for one to (partially) land on the somehow safe and familiar grounds of NRQM.
Can we recover a position operator or a local number operator in the non relativistic regime?
These questions are based on the intuition that relativistic QFT, being the best known theory that combines quantum theory with special relativity, should `contain' its predecessor NRQM.
This expectation is because if QFT is indeed more fundamental than NRQM, then it needs to account for the experimentally-verified predictions of NRQM.
However, one needs to specify what is meant by the containment of NRQM within QFT, since it is far from straightforward which features of the predecessor theory can be retrieved.
Such a retentionist view is further obscured by the fact that different theories suggest different ontological commitments, which means that part of the fundamental ontology of a theory is dropped once the `next' theory is established.
Here particle ontology is dropped in the case of relativistic QFT succeeding NRQM.
This means that even if one is able to identify elements of the mathematical description of the previous theory, here a wavefunction or a position operator, the interpretation of these objects will be essentially different.
The main challenge is to identify these differences, even in cases that some mathematical elements are relatively easy to retrieve.
As we find, the recovery of NRQM from relativistic QFT is only approximate, and we will discuss how it differs from standard NRQM.
For a philosophical perspective see \cite{Wallace2001,myrvold2015wavefunction, Benjamin}, and for recent developments \cite{Padmanabhan2018}.

Another undertaking of this paper is to clarify whether \updated{ground state entanglement} is a feature of the relativistic regime of \updated{relativistic} QFT.
At first, this may seem unrelated to the issue of particle localizability, but the Reeh-Schlieder theorem provides this connection through the intuition expounded in \cite{Redhead1995,VazquezEtal2014}.
\updated{Also, the presence of entanglement in the ground state of a relativistic QFT is starkly manifested in the Unruh effect.
Since ground state entanglement is a necessary (although not sufficient) ingredient for both the Reeh-Schlieder theorem and the Unruh effect, these features would subside in the non-relativistic regime if the ground state entanglement were to vanish.}
\updated{Of course, there are many non-relativistic systems (e.g., in condensed matter) with ground states exhibiting entanglement.
The focus of the investigation here is whether ground state entanglement in a relativistic QFT depends on its relativistic nature.
Here we aim to address this by examining whether non-zero measures of this entanglement persist in the non-relativistic regime of the relativistic QFT.}

In Section~\ref{sec:background}, we review the different Hilbert spaces that one can associate to a free QFT and how they relate to each other.
The continuous tensor product that can naturally accomodate the infinite degrees of freedom of the QFT is suitably defined as a Fock space, both in momentum and in position space (the global and local Fock spaces, respectively).
We derive the Bogoliubov transformation between the local and the global degrees of freedom, which allows us to concretely compare two different localization schemes in QFT, which we refer to as `standard' and `non-relativistic'.
In Section~\ref{sec:NR_limit}, we describe our implementation of the non-relativistic approximation by means of a bandlimited theory.
We identify the subspace of the QFT Hilbert space where the approximation holds, in order determine whether the non-relativistic localization scheme faithfully represents the local degrees of freedom in this subspace.
As we find, the answer is negative, but yet it is by means of this scheme that we can recover features of NRQM in this subspace, as described in Section~\ref{sec:localizability}.
The cutoff we need to recruit to implement the non-relativistic approximation non-trivially affects the local degrees of freedom and the entanglement that they share.
In Section~\ref{sec:entanglement}, we quantify the remaining entanglement between these local degrees of freedom.
These considerations reduce to examining the expression for the $\beta$-coefficients in the Bogoliubov transformation between the local and global annihilation and creation operators.
Here we consider the non-relativistic dispersion relation as an approximation to the relativistic one, but many of the conclusions would qualitatively hold for quantum field theories with more general dispersion relations, as occur with condensed matter systems, sonic analogues \cite{Unruh1981,Unruh1995}, and quantum gravity inspired effective Planck-scale corrections \cite{AmelinoCameliaEtal1998,Gambini-Pullin1999,Magueijo-Smolin2002,Myers-Pospelov2003,GirelliEtal2007}.

\section{Background: Local and global Fock spaces}\label{sec:background}

In this section, we review two quantization schemes (called local and global) for a massive real Klein-Gordon field, as well as how these two schemes are related to each other.
This will serve to establish notation and to emphasize aspects which will be important for the subsequent discussion.

\subsection{The Hilbert space zoo of QFT}

The Hilbert space that is most commonly attributed to a free QFT is a Fock space.
If we aspire to describe a free QFT in terms of particles, and demand that particles are entities that can be counted by an observable number operator, then the full Hilbert space of the QFT should decompose into a direct sum of fixed particle number subspaces.
Explicitly, if the single-particle subspace is described by a Hilbert space, $\mathcal{H}$, then the corresponding Fock space is:
\begin{equation}
  \mathcal{F}[\mathcal{H}] := \bigoplus_{n=0}^\infty ( \mathcal{H}^{\otimes n} )_{S,A} = \mathds{C} \oplus \mathcal{H} \oplus ( \mathcal{H}^{\otimes 2} )_{S,A} \oplus \cdots , \label{fock}
\end{equation}
where the subscripts $S$ and $A$ denote symmetrization and anti-symmetrization of the tensor product, for bosonic or fermionic particles respectively.
Given a particular QFT, one can build many different Fock spaces to use as the state space.
The number operators associated with each of these Fock spaces are used to count different entities.
Furthermore, it is a unique feature of quantum theories of infinite degrees of freedom that there exist unitarily inequivalent representations of the canonical commutation relations, due to the absence of an analogue of the Stone-von Neumann theorem \cite{Haag1992,Ruetsche2011}.
As a result, some of these Fock spaces may be unitarily inequivalent\footnote{When referring to unitary inequivalence of two Fock spaces, indeed we do not mean that the two Hilbert spaces are not isomorphic (e.g., any two separable Hilbert spaces are isomorphic), but rather that the representations of the algebra generated by a set of operators satisfying the canonical commutation relations cannot be intertwined using a unitary operator.}, and the associated notions of counting will be incommensurable.

The typical starting point for concretely constructing the state space of a QFT is through recruiting the analogy with an infinite collection of harmonic oscillators, either in position or momentum space.
The corresponding Hilbert space is then formally defined as the infinite tensor product over this collection of harmonic oscillators.
The most naive quantization procedure for a massive real Klein-Gordon field begins by viewing the Hamiltonian written in position space as a collection of harmonic oscillators, labelled by the position $\boldsymbol{x}$ (in $n$ spatial dimensions) and coupled through the spatial derivative,
\begin{equation}\label{eq:ham}
  H = \frac12 \int d\boldsymbol{x} \hspace{1mm} [ c^2 \Pi(\boldsymbol{x})^2 + ( \boldsymbol{\nabla} \Phi(\boldsymbol{x}) )^2 + k_c^2 \Phi(\boldsymbol{x})^2 ],
\end{equation}
where $k_c := mc/\hbar$ is the wavenumber associated with the Compton scale.
One can then proceed to construct a Fock space for each $\boldsymbol{x}$ by defining local annihilation (and corresponding creation) operators as,
\begin{equation}\label{eq:bx_defn}
  \hat{b}_{\boldsymbol{x}} := \sqrt{\frac{m}{2\hbar^2}} \hat{\Phi}(\boldsymbol{x}) + \frac{i}{\sqrt{2m}} \hat{\Pi}(\boldsymbol{x}).
\end{equation}
These annihilation and creation operators are deemed local because they are labeled with the same point, $\boldsymbol{x}$, as the field operators, but also because they diagonalize the ``uncoupled'' terms in the Hamiltonian, i.e.,
\begin{equation}
  \frac12 \int d\boldsymbol{x} \hspace{1mm} [ c^2 \hat{\Pi}(\boldsymbol{x})^2 + k_c^2 \hat{\Phi}(\boldsymbol{x})^2 ] = \frac12 \int d\boldsymbol{x} \hspace{1mm} mc^2 ( \hat{b}_{\boldsymbol{x}}^\dagger \hat{b}_{\boldsymbol{x}} + \hat{b}_{\boldsymbol{x}} \hat{b}_{\boldsymbol{x}}^\dagger ).
\end{equation}
We will consider the Fock space associated with the degree of freedom at the point $\boldsymbol{x}$ to be that generated by $\hat{b}_{\boldsymbol{x}}$ and $\hat{b}_{\boldsymbol{x}}^\dagger$, explicitly given by $\mathcal{H}_{\boldsymbol{x}} := L_2(\mathds{R},d\Phi(\boldsymbol{x}))$.
We then formally take a continuous tensor product over these Fock spaces to construct a Hilbert space for the full QFT, $\otimes_{\boldsymbol{x}} \mathcal{H}_{\boldsymbol{x}}$.
The most common means of defining such a continuous tensor product leads to a non-separable Hilbert space \cite{vonNeumann1939,Wald1994,Streater-Wightman1964}.
Furthermore, it is unclear how to connect this construction of the full Hilbert space to the general form of a Fock space as defined above.
That is, can one write $\otimes_{\boldsymbol{x}} \mathcal{H}_{\boldsymbol{x}} \cong \mathcal{F}[\mathcal{H}]$ for some $\mathcal{H}$?
It turns out that such a description is possible using an alternative definition of the continuous tensor product, which also leads to a separable Hilbert space.
This alternative definition uses a construction called the \emph{exponential Hilbert space} \cite{Streater1969,Isham-Linden1995,Isham-Linden-Savvidou-Schreckenberg1998,Klauder1970}, which is applicable if $\mathcal{H}_{\boldsymbol{x}}$ is a Fock space for every $\boldsymbol{x}$, as is the case in the QFT considered here.
Following \cite{Isham-Linden1995,Isham-Linden-Savvidou-Schreckenberg1998}, we make the identification,
\begin{equation}
  \otimes_{\boldsymbol{x}} \mathcal{H}_{\boldsymbol{x}} = \otimes_{\boldsymbol{x}} L_2(\mathds{R},d\Phi(\boldsymbol{x})) \cong \mathcal{F}[L_2(\mathds{R}^n,d\boldsymbol{x})] =: \mathcal{F}_L.
\end{equation}
Thus, we can indeed identify the full Hilbert space of the QFT with a Fock space, which we will call the \emph{local Fock space}, $\mathcal{F}_L$.
Correspondingly, we will denote the total local number operator by $\hat{N}_L := \int d\boldsymbol{x} \hspace{1mm} \hat{b}_{\boldsymbol{x}}^\dagger \hat{b}_{\boldsymbol{x}}$, and the local vacuum as $\ket{0}_L := \otimes_{\boldsymbol{x}} \ket{0}_{\boldsymbol{x}}$ (i.e., $\hat{b}_{\boldsymbol{x}} \ket{0}_L = 0$ for all $\boldsymbol{x}$).
It may seem that our ability to define these objects is in contradiction with the Reeh-Schlieder theorem, which implies the non-existence of local number operators.
However, we will see that this is not the case, as the Reeh-Schlieder theorem is formulated in the context of a different Fock space construction.

Distinct from the local quantization, the most popular quantization scheme, as seen in most particle physics textbooks, is to quantize the normal modes of the free Hamiltonian of the theory.
In flat spacetime, the normal modes for the free Klein-Gordon theory are simply plane waves.
In Fourier space, the Hamiltonian is:
\begin{equation}
  H = \frac12 \int \frac{d\boldsymbol{k}}{(2\pi)^n} \left[ c^2 |\Pi_{\boldsymbol{k}}|^2 + \frac{\omega_{\boldsymbol{k}}^2}{c^2} |\Phi_{\boldsymbol{k}}|^2 \right],
\end{equation}
with $\omega_{\boldsymbol{k}} := c \sqrt{\boldsymbol{k}^2 + k_c^2}$.
The Fock space for mode $\boldsymbol{k}$ is constructed with annihilation (and corresponding creation) operators,
\begin{equation}
  \hat{a}_{\boldsymbol{k}} := \sqrt{\frac{\omega_{\boldsymbol{k}}}{2\hbar c^2}} \hat{\Phi}_{\boldsymbol{k}} + i \sqrt{\frac{c^2}{2\hbar \omega_{\boldsymbol{k}}}} \hat{\Pi}_{\boldsymbol{k}},
\end{equation}
and can be written as\footnote{Note that we are using a slight abuse of notation: since we are only considering real-valued fields, the functions in this space must satisfy the constraint $\Phi_{\boldsymbol{k}}^\ast = \Phi_{-\boldsymbol{k}}$.
However, this technical point will not be of concern in the following discussion as it is accounted for implicitly by writing the field operator as $\hat{\Phi}_{\boldsymbol{k}} = \sqrt{\frac{\hbar c^2}{2\omega_{\boldsymbol{k}}}} ( \hat{a}_{\boldsymbol{k}} + \hat{a}_{-\boldsymbol{k}}^\dagger )$.} $\mathcal{H}_{\boldsymbol{k}} := L_2(\mathds{C},d\Phi_{\boldsymbol{k}})$.
In a similar manner to the above identification for the continuous tensor product with the local Fock space, we have
\begin{equation}
  \otimes_{\boldsymbol{k}} \mathcal{H}_{\boldsymbol{k}} = \otimes_{\boldsymbol{k}} L_2(\mathds{C},d\Phi_{\boldsymbol{k}}) \cong \mathcal{F}[L_2(\mathds{R}^n,d\boldsymbol{k})] =: \mathcal{F}_G.
\end{equation}
We will refer to this Fock space as the \emph{global Fock space}, $\mathcal{F}_G$, in contrast with the local Fock space, $\mathcal{F}_L$, defined above.
The corresponding total global number operator is $\hat{N}_G := \int \frac{d\boldsymbol{k}}{(2\pi)^n} \hat{a}_{\boldsymbol{k}}^\dagger \hat{a}_{\boldsymbol{k}}$, and the global vacuum state is defined to be $\ket{0}_G := \otimes_{\boldsymbol{k}} \ket{0}_{\boldsymbol{k}}$ (i.e., $\hat{a}_{\boldsymbol{k}} \ket{0}_G = 0$ for all $\boldsymbol{k}$).
Of course, the global number operator commutes with the Hamiltonian and the global vacuum is its ground state.

One purpose of choosing the normal mode (or global) quantization scheme is that if we insist on particles being entities that can be counted, then they should also have an element of persistence.
(Of course, this will not be appropriate for interacting theories, however the localizability issues we are considering here are present even in the case of free theories.
That being said, the discussion may also be relevant for the free theories one identifies with the asymptotic future and past of interacting theories, but we will not pursue this further here.)
A means of formalizing the requirement of persistence is to demand that the particle number be conserved in time, i.e., the number operator of the Fock space should commute with the Hamiltonian.
For the free Klein-Gordon theory, this is only true of the global number operator.
Another appeal of the global quantization is that the entities created with $\hat{a}_{\boldsymbol{k}}^\dagger$ exhibit the appropriate relativistic dispersion relation between mass, momentum, and energy: $\hat{H} \left( \hat{a}_{\boldsymbol{k}}^\dagger \ket{0}_G \right) = \sqrt{ c^2 \hbar^2 \boldsymbol{k}^2 + m^2 c^4 } \left( \hat{a}_{\boldsymbol{k}}^\dagger \ket{0}_G \right)$. 

Of course, one would not expect an excitation of the form $\hat{a}_{\boldsymbol{k}}^\dagger \ket{0}_G$ to exhibit any sensible notion of localizability, since the normal modes of the Klein-Gordon field are supported everywhere in space.
A simple remedy would seem to be to define a wavepacket state of the form $\ket{\Psi} = \int \frac{d\boldsymbol{k}}{(2\pi)^n} \tilde{f}(\boldsymbol{k}) \hat{a}_{\boldsymbol{k}}^\dagger \ket{0}_G$, which lies in the single-particle subspace of the global Fock space and exhibits the appropriate dispersion relation in each branch of the momentum superposition.
It seems that one could then choose a wavepacket, $\tilde{f}(\boldsymbol{k})$, so that its Fourier transform is arbitrarily well localized in space, while retaining the appealing features of the global quantization.
However, this is only illusory, for it turns out that it is not appropriate to characterize the localizability of this wavepacket state with the Fourier transform of $\tilde{f}(\boldsymbol{k})$.
The reasoning for this stems from how this Fourier-transformed space relates to the local operators associated with the local Fock space.
The remainder of this section is devoted to elucidating this relationship.

\subsection{Relating the local and global Fock spaces}

The first step will be to examine how the local and global Fock spaces relate to each other.
The relationship is most clearly seen through the Bogoliubov transformation between the creation and annihilation operators of the two quantization schemes:
\begin{equation}\label{eq:xk_bbv}
  \hat{b}_{\boldsymbol{x}} = \int \frac{d\boldsymbol{k}}{(2\pi)^n} e^{i \boldsymbol{k} \cdot \boldsymbol{x}} \left[ c_+(\boldsymbol{k}) \hat{a}_{\boldsymbol{k}} - c_-(\boldsymbol{k}) \hat{a}_{-\boldsymbol{k}}^\dagger \right],
\end{equation}
where
\begin{equation}
  c_\pm(\boldsymbol{k}) := \frac12 \left[ (1+(\boldsymbol{k}/k_c)^2)^{1/4} \pm (1+(\boldsymbol{k}/k_c)^2)^{-1/4} \right].
\end{equation}
The fact that there is mixing between the local and global annihilation and creation operators demonstrates that the counting is different in the two Fock spaces $\mathcal{F}_L$ and $\mathcal{F}_G$.
Indeed, the assertion of the Reeh-Schlieder theorem is that there cannot be local number operators counting the particles of the global Fock space, $\mathcal{F}_G$.
Moreover, one can show that the two representations are unitarily inequivalent (see Appendix~\ref{apdx:bbv}).
Hence these two notions of counting are incommensurable in the sense that the expectation value of $\hat{N}_L$ on any state of $\mathcal{F}_G$ diverges (and vice-versa).

That the local and global Fock spaces are different clearly demonstrates that the ground state of the Hamiltonian is not the local vacuum,
\begin{equation}
  \ket{0}_G = \otimes_{\boldsymbol{k}} \ket{0}_{\boldsymbol{k}} \neq \ket{0}_L = \otimes_{\boldsymbol{x}} \ket{0}_{\boldsymbol{x}}
\end{equation}
(in fact, they do not even lie in the same space).
Intuitively, this seems to indicate that the local degrees of freedom are entangled in the ground state (global vacuum).
As discussed in \cite{Redhead1995}, it is precisely this entanglement which is responsible for preventing the existence of local number operators counting the particles of $\mathcal{F}_G$.
We will reconsider this ground state entanglement between local degrees of freedom in Section~\ref{sec:entanglement}, after making the non-relativistic approximation of the field theory.

\subsection{Two localization schemes in QFT}\label{subsec:two_schemes}

In the above quantization schemes, we began by writing degrees of freedom labeled by points in either position or momentum space, and then formally constructed the Hilbert space by taking a tensor product over these labels.
We then proceeded to identify this full space with a corresponding Fock space.
The Fock space picture is useful when discussing particle notions, but we do not want to abandon the tensor product description altogether, as this is how one can identify subsystems of the field, which is useful, e.g., if one is interested in entanglement.

There are many ways one can choose such a tensor product structure and arrive at the same Fock space.
For example, suppose we have a basis, $\{ f_i(\boldsymbol{x}) \}_i$, for the space of classical field configurations.
Then we can define coefficients of the classical fields as $\Phi_i := \int d\boldsymbol{x} f_i(\boldsymbol{x}) \Phi(\boldsymbol{x})$ and construct the full Hilbert space by again using the analogy with a collection of harmonic oscillators, but now labeled by the elements in the basis.
Explicitly, we can formally write the full Hilbert space as $\otimes_i L_2(\mathds{R},d\Phi_i)$, where each of the sectors labeled with $i$ is constructed as a Fock space with the annihilation operators $\hat{b}_i := \sqrt{\tfrac{m}{2\hbar^2}} \hat{\Phi}_i + \tfrac{i}{\sqrt{2m}} \hat{\Pi}_i = \int d\boldsymbol{x} f_i(\boldsymbol{x}) \hat{b}_{\boldsymbol{x}}$.
Note that we can reconstruct the continuously labeled field and annihilation operators by $\hat{\Phi}(\boldsymbol{x}) = \sum_i \hat{\Phi}_i f_i(\boldsymbol{x})$ and $\hat{b}_{\boldsymbol{x}} = \sum_i \hat{b}_i f_i(\boldsymbol{x})$.
Since the transformation between the two sets of annihilation operators does not mix with the creation operators, then clearly this new set of annihilation and creation operators generate the local Fock space, i.e., $\otimes_i L_2(\mathds{R},d\Phi_i) \cong \mathcal{F}[L_2(\mathds{R}^n,d\boldsymbol{x})]$.

We can view the expression of the annihilation and creation operators in the basis $\{ f_i(\boldsymbol{x}) \}_i$ as simply a change of basis in the single-particle subspace.
Conversely, every change of basis in the single-particle subspace of the Fock space corresponds to a different tensor product decomposition of the full Hilbert space.
In this sense, changing a basis in this subspace can be seen as a rearrangement of the degrees of freedom, or a different infinite collection of harmonic oscillators through which we are describing the quantum field theory.
However, these changes of bases do not exhaust all of the possible tensor product structures (hence subsystem decompositions) which are available, since they cannot change the Fock space.
We have already seen this above, where we took a tensor product over normal modes and arrived at a different Fock space.

Of course, we can perform these changes of basis in the global Fock space as well.
For instance, one could use a plane wave basis to define the Fourier transforms of the global annihilation and creation operators,
\begin{equation}\label{eq:ay_defn}
  \hat{a}_{\boldsymbol{y}} := \int \frac{d\boldsymbol{k}}{(2\pi)^n} e^{i \boldsymbol{k} \cdot \boldsymbol{y}} \hat{a}_{\boldsymbol{k}}.
\end{equation}
Notice that these new set of annihilation and creation operators characterize the Fourier-transformed wavepacket states we had discussed above, i.e.,
\begin{equation}
  \ket{\Psi} = \int \frac{d\boldsymbol{k}}{(2\pi)^n} \tilde{f}(\boldsymbol{k}) \hat{a}_{\boldsymbol{k}}^\dagger \ket{0}_G = \int d\boldsymbol{y} f(\boldsymbol{y}) \hat{a}_{\boldsymbol{y}}^\dagger \ket{0}_G,
\end{equation}
where $\tilde{f}(\boldsymbol{k})$ is the Fourier transform of $f(\boldsymbol{y})$.
Above we considered the possibility that one could choose an arbitrarily well localized wavepacket, $f(\boldsymbol{y})$, to obtain a localized state of the field theory which exhibits the desirable features of the global quantization (such as persistence in time and obeying the appropriate relativistic dispersion relation).
However, the issue in doing so is that the function $f(\boldsymbol{y})$ does not appropriately characterize the localizability of the wavepacket in space.

To see this, first we note that in the above wavepacket state, the amplitude of the function $f$ at $\boldsymbol{y}$ determines the amplitude of the excitation created with $\hat{a}_{\boldsymbol{y}}^\dagger$.
Hence, if this function is highly peaked around a particular point $\boldsymbol{y}$, we can consider this excitation to be created ``at $\boldsymbol{y}$''.
However, perhaps counter to naive expectations, this $\boldsymbol{y}$ does not correspond to a point in space.
We can see this by first combining the definition \eqref{eq:ay_defn} and the Bogoliubov transformation \eqref{eq:xk_bbv}:
\begin{equation}\label{eq:yx_bbv}
  \hat{a}_{\boldsymbol{y}} = \int d\boldsymbol{x} [ F_+(\boldsymbol{y}-\boldsymbol{x}) \hat{b}_{\boldsymbol{x}} + F_-(\boldsymbol{y}-\boldsymbol{x}) \hat{b}_{\boldsymbol{x}}^\dagger ],
\end{equation}
where $F_\pm(\boldsymbol{y}-\boldsymbol{x}) := \int \frac{d\boldsymbol{k}}{(2\pi)^n} e^{i \boldsymbol{k} \cdot (\boldsymbol{y}-\boldsymbol{x})} c_\pm(\boldsymbol{k})$.
We stated above that the local annihilation and creation operators, $\hat{b}_{\boldsymbol{x}}$ and $\hat{b}_{\boldsymbol{x}}^\dagger$, are labeled by points in space since they are directly related to the local field operators.
Therefore, one sees from this Bogoliubov transformation that the operators $\hat{a}_{\boldsymbol{y}}$ and $\hat{a}_{\boldsymbol{y}}^\dagger$ act non-locally.
The degree of non-locality is governed by the integral kernels $F_\pm$, which decay asymptotically as $F_\pm \sim e^{-k_c |\boldsymbol{y}-\boldsymbol{x}|}$.
Hence the non-locality of the operators $\hat{a}_{\boldsymbol{y}}$ is suppressed exponentially at distances much larger than the Compton wavelength of the field.

Note that from the definition of $\hat{a}_{\boldsymbol{y}}$, it follows that $\ket{0}_G = \otimes_{\boldsymbol{y}} \ket{0}_{\boldsymbol{y}}$, since the two sets of operators generate the same Fock space (and, in particular, annihilate the same vacuum).
Therefore, there is clearly no vacuum entanglement between the $\boldsymbol{y}$ degrees of freedom in the ground state of the field theory.

Hence, we arrive at two different localization schemes for our QFT: that defined in \eqref{eq:bx_defn} using the local operators, which we will refer to as the `standard' localization scheme, and the other defined in \eqref{eq:ay_defn} using the Fourier-transformed global operators.
Features of the two localization schemes have been investigated in \cite{Halvorson-Clifton2002,Piazza-Costa2007}.
In the literature, the operators $\hat{a}_{\boldsymbol{y}}$ and $\hat{a}_{\boldsymbol{y}}^\dagger$ are commonly introduced as non-relativistic field operators (see, e.g., \cite{Anastopoulos-Hu2014,Piazza-Costa2007,Bjorken-Drell1964}).
\updated{A summary of the main features of the two localization schemes is provided in Table \ref{tab:two_schemes}.}

\begin{table}[h!]
\updated{
\caption{\label{tab:two_schemes}Summary of the localization schemes.}
\begin{center}
\begin{tabular}{ | c | c | }
    \hline
    \textbf{Non-relativistic}: $\hat{a}_{\boldsymbol{y}}, \hat{a}_{\boldsymbol{y}}^\dagger$  &  \textbf{Local}: $\hat{b}_{\boldsymbol{x}}, \hat{b}_{\boldsymbol{x}}^\dagger$ \\
    \hline
    Fourier transforms of $\hat{a}_{\boldsymbol{k}}, \hat{a}_{\boldsymbol{k}}^\dagger$  &  Bogoliubov mixing of $\hat{a}_{\boldsymbol{k}}, \hat{a}_{\boldsymbol{k}}^\dagger$ \\
    \hline
    non-locally related to $\hat{\Phi}(\boldsymbol{x}), \hat{\Pi}(\boldsymbol{x})$  &  locally related to $\hat{\Phi}(\boldsymbol{x}), \hat{\Pi}(\boldsymbol{x})$ \\
    \hline
    act in $\mathcal{F}_G$  &  act in $\mathcal{F}_L$, not $\mathcal{F}_G$ \\
    \hline
      $\ket{0}_G = \otimes_{\boldsymbol{y}} \ket{0}_{\boldsymbol{y}}$  &  $\ket{0}_G \neq \otimes_{\boldsymbol{x}} \ket{0}_{\boldsymbol{x}}$ \\
    \hline
\end{tabular}
\end{center}
}
\end{table}

One might expect that these operators coincide with the local annihilation and creation operators in the non-relativistic regime.
This would also be consistent with the observation that the degree of non-locality of the integral kernels $F_\pm$ in \eqref{eq:yx_bbv} decays exponentially beyond the Compton scale, and hence this non-locality may become insignificant in this regime.
One would then be able to use these non-relativistic field operators to define localizable particle states exhibiting the features of the global quantization.
Furthermore, if the two localization schemes were to coincide in the limit, the local and global vacua would also coincide, indicating that any entanglement between the local degrees of freedom would be unobservable in this limit.
Intuitively, this would have the consequence of lifting obstructions to particle localizability due to the Reeh-Schlieder theorem.

All of this would be a simple and consistent story, if only it were true.
It turns out the reasons the `non-relativistic' scheme is appropriate in the non-relativistic regime (thus justifying the terminology) is more subtle, as we will now discuss.

\section{Aspects of the non-relativistic approximation}\label{sec:NR_limit}

In this section we will describe the manner in which we will implement the non-relativistic approximation for the Klein-Gordon theory described above.
In the literature, there are different approaches for making this approximation; one can find methods using the WKB approximation or similar techniques \cite{Proca1938,Bjorken-Drell1964,Kiefer-Singh1991}, others using a group theoretic perspective \cite{Weinberg1995,LevyLeblond1967}, and yet others using a renormalization perspective \cite{daSilva2001,Caswell-Lepage1986,Lepage1989}.
Often in these approaches, one finds either the implicit or explicit assumption of an ultraviolet cutoff that can suitably impose the ``small momentum'' condition.
Here we will be careful to be explicit about this assumption, as it will have implications for our following investigation into localizability and vacuum entanglement.

\subsection{Requirement of an ultraviolet cutoff}

For a single classical particle, the non-relativistic approximation can be stated in terms of an expansion of the energy \emph{to second order} in $|\boldsymbol{p}|/mc$, i.e., $E = \sqrt{ m^2 c^4 + \boldsymbol{p}^2 c^2 } \approx mc^2 + \boldsymbol{p}^2 / 2m$.
For a quantum particle, we use the de Broglie relation $\boldsymbol{p} = \hbar \boldsymbol{k}$ to conclude that the appropriate expansion is to second order in $|\boldsymbol{k}|/k_c$.

There are perhaps different means through which one could go about formally implementing this approximation.
For a single-particle wavepacket of the form $\int \frac{d\boldsymbol{k}}{(2\pi)^n} \psi(\boldsymbol{k}) \hat{a}_{\boldsymbol{k}}^\dagger \ket{0}_G$, one may consider declaring that such particles are `slow' if the expectation value and the variance of the momentum is small.
For example, this could be achieved by restricting the set of allowable wavepackets to those with suitably quick decay at large momenta.
However, this space fails to be a closed linear space, which is a requirement if we want the resulting space to form a Hilbert space.
Because this space of wavepackets will ultimately be the state space we use for NRQM, it seems that this requirement is appropriate so that we recover important structural features of NRQM, such as the spectral theorem for observables restricted to this subspace.
The cost of enforcing this requirement is to impose a cutoff on the set of allowable wavenumbers, i.e., the support of these wavepackets should be restricted to $|\boldsymbol{k}| < \Lambda$, where $\Lambda$ is a cutoff such that $\Lambda/k_c \ll 1$.
The corresponding single-particle Hilbert space (called \emph{bandlimited wavefunctions}) will be denoted $B(\Lambda)$.
Imposing a sharp cutoff of this kind also confines us to a set of states where particle pair creation cannot occur (in an interacting theory).
Since particle creation and annihilation is a fundamental difference between QFT and NRQM, this seems an appropriate restriction, as a softer cutoff would always provide a non-zero amplitude for these processes.
Notice also that imposing such a cutoff picks out a preferred frame (that is, we break the Lorentz-invariance of the theory at this stage), but of course there should not be a frame-independent notion of `slow'.

It is straightforward to extend this to multiparticle states.
If we consider a two-particle state $\int \frac{d\boldsymbol{k}_1}{(2\pi)^n} \frac{d\boldsymbol{k}_2}{(2\pi)^n} \psi(\boldsymbol{k}_1,\boldsymbol{k}_2) \hat{a}_{\boldsymbol{k}_1}^\dagger \hat{a}_{\boldsymbol{k}_2}^\dagger \ket{0}_G$, then the appropriate restriction would be to limit the support of the wavepacket in both variables to $|\boldsymbol{k}_1|, |\boldsymbol{k}_2| < \Lambda$.
This is because we want the momenta of \emph{each} of the particles to be small, and not the \emph{total} momentum, for example.
Hence, the two-particle subspace should be identified with $(B(\Lambda)^{\otimes 2})_S$, and similarly for higher particle-number subspaces.
Overall, we arrive at a Fock space constructed with symmetrized tensor products of $B(\Lambda)$, namely $\mathcal{F}[B(\Lambda)]$.
This is a subspace of the global Fock space, since
\begin{equation}
  B(\Lambda) \subset L_2(\mathds{R}^n,d\boldsymbol{k}) \implies \mathcal{F}[B(\Lambda)] \subset \mathcal{F}[L_2(\mathds{R}^n,d\boldsymbol{k})] = \mathcal{F}_G.
\end{equation}

One can also think of this bandlimited Fock space as obtained by removing (tracing out) the set of degrees of freedom associated with wavenumbers above the cutoff, i.e.,
\begin{equation}
  \mathcal{F}[B(\Lambda)] \cong \otimes_{|\boldsymbol{k}|<\Lambda} L_2(\mathds{C},d\Phi_{\boldsymbol{k}}) \subset \mathcal{F}[L_2(\mathds{R}^n,d\boldsymbol{k})] \cong \otimes_{\boldsymbol{k}} L_2(\mathds{C},d\Phi_{\boldsymbol{k}}).
\end{equation}
As has been studied in \cite{Pye-Donnelly-Kempf2015}, bandlimitation has a non-trivial effect on the structure of local degrees of freedom and entanglement of the field.
We will return to explicitly discuss these effects in Subsection~\ref{subsec:sampling} and Section~\ref{sec:entanglement} (respectively).
Regardless of whether the cutoff is introduced for fundamental or for phenomenological reasons, all of the implications of bandlimitation need to be taken into account.
In \cite{Pye-Donnelly-Kempf2015}, quantum gravity considerations motivated introducing a cutoff at the Planck scale.
For the purposes of the non-relativistic approximation, the size of the cutoff is rather set by the Compton scale.
An operational means through which one could motivate introducing such a cutoff is through an interface with a probing system which couples only to this subset of modes.
This could occur, for example, in a detector model which couples to the field via a bandlimited smearing function, intuitively corresponding to a large detector (compared to the Compton wavelength of the field).
We can consistently describe the physics restricted to $\mathcal{F}[B(\Lambda)] \cong \otimes_{\boldsymbol{k}<\Lambda} L_2(\mathds{C},d\Phi_{\boldsymbol{k}})$, since in the free theory each $\boldsymbol{k}$ sector is decoupled.

We began this discussion by considering wavepackets formed with the global creation operators and seemingly abandoned the local Fock space associated with the local annihilation and creation operators $\hat{b}_{\boldsymbol{x}}$ and $\hat{b}_{\boldsymbol{x}}^\dagger$.
We choose to focus on the global Fock space as it is these particle excitations which exhibit the correct dispersion relation and are preserved in time.
If we are to hope to end up with a non-relativistic theory for fixed particle number states, with the Schr\"odinger equation as the equation of motion for these wavepackets, then we must choose to focus on states residing in the global Fock space.
However, we will return to consider the relevance of the local operators in the non-relativistic regime.

The bandlimited Fock space, $\mathcal{F}[B(\Lambda)]$, is the space from which we will draw non-relativistic wavefunctions.
Because this is a subspace of the global Fock space, it is straightforward to define the restriction of operators from the field theory to this subspace.
For example, the total momentum operator becomes
\begin{equation}
  \boldsymbol{\hat{P}} \big|_{\mathcal{F}[B(\Lambda)]} = \int_{|\boldsymbol{k}|<\Lambda} \frac{d\boldsymbol{k}}{(2\pi)^n} \hbar \boldsymbol{k} \hat{a}_{\boldsymbol{k}}^\dagger \hat{a}_{\boldsymbol{k}},
\end{equation}
which we see is simply a restriction of the integration range of the $\boldsymbol{k}$ values.

\subsection{Operator approximations}

Once an operator written as an integral over wavevectors has been bandlimited, one is able to expand the integrand in powers of $|\boldsymbol{k}|/k_c \leq \Lambda/k_c \ll 1$.
The non-relativistic approximation entails keeping terms up to second order in this ratio.
More concretely, let us consider again the Bogoliubov transformation \eqref{eq:xk_bbv}.
After introducting a cutoff, we can write the bandlimited local annihilation operator using the (inverse) Bogoliubov transformation:
\begin{equation}
  \hat{b}_{\boldsymbol{x}} = \int_{|\boldsymbol{k}|<\Lambda} \frac{d\boldsymbol{k}}{(2\pi)^n} e^{i \boldsymbol{k} \cdot \boldsymbol{x}} \left[ c_+(\boldsymbol{k}) \hat{a}_{\boldsymbol{k}} - c_-(\boldsymbol{k}) \hat{a}_{-\boldsymbol{k}}^\dagger \right],
\end{equation}
with $c_\pm(\boldsymbol{k}) := \frac12 \left[ (1+(\boldsymbol{k}/k_c)^2)^{1/4} \pm (1+(\boldsymbol{k}/k_c)^2)^{-1/4} \right]$.
Since $c_\pm(\boldsymbol{k})$ are analytic at 0, we can expand these in a Maclaurin series in $|\boldsymbol{k}|/k_c < \Lambda/k_c \ll 1$.
To second order, these become $c_+(\boldsymbol{k}) \sim 1$ and $c_-(\boldsymbol{k}) \sim \tfrac14 (\boldsymbol{k}/k_c)^2$.
Keeping up to second order terms, we can insert these into the above Bogoliubov transformation to get
\begin{equation}\label{eq:bx_NR}
  \hat{b}_{\boldsymbol{x}} = \int_{|\boldsymbol{k}|<\Lambda} \frac{d\boldsymbol{k}}{(2\pi)^n} e^{i \boldsymbol{k} \cdot \boldsymbol{x}} \left[ \hat{a}_{\boldsymbol{k}} - \frac14 (\boldsymbol{k}/k_c)^2 \hat{a}_{-\boldsymbol{k}}^\dagger \right].
\end{equation}
Hence we obtain an operator that is equivalent to the original operator $\hat{b}_{\boldsymbol{x}}$ for modes below the bandlimit (to second order in $\Lambda/k_c$).
In the following, we will interpret this as the action of $\hat{b}_{\boldsymbol{x}}$ on the bandlimited subspace in the non-relativistic regime.
The extension of this procedure to other operators written as Fourier integrals is straightforward.

\section{Localizability under the non-relativistic approximation}\label{sec:localizability}

In this section we show that the two localization schemes do not coincide in the non-relativistic regime, and we demonstrate which should serve to salvage NRQM from QFT under the non-relativistic approximation.
We describe the sense in which we recover some of the features of standard NRQM, e.g., what can play the role of a wavefunction and a position operator. 
Despite obtaining analogues of these objects, we find that there remain limitations to the localizability properties of these wavefunctions.
We will also discuss the form that the spatial degrees of freedom of the field theory and the wavefunctions take after the non-relativistic approximation.

\subsection{Localization schemes in the non-relativistic regime}\label{subsec:two_schemes_NR}

In Subsection~\ref{subsec:two_schemes}, we presented both the standard and non-relativistic schemes for characterizing localizability of particle states in a quantum field theory.
We also alluded to the possibility that the non-locality of the non-relativistic operators, inferred from the Bogoliubov transformation \eqref{eq:yx_bbv}, may disappear in the non-relativistic regime of the quantum field theory.
In this way, the non-relativistic localization scheme would faithfully represent the local degrees of freedom in this regime.
However, using the tools outlined in Section~\ref{sec:NR_limit}, we will now demonstrate that the two schemes do not coincide in this regime.

The Bogoliubov transformation \eqref{eq:yx_bbv} between the local and non-relativistic operators will be the relation of central importance to this demonstration.
Introducing a cutoff $|\boldsymbol{k}|<\Lambda$ and expanding to second order in $\Lambda/k_c$, we find
\begin{equation}\label{eq:yx_bbv_NR}
  \hat{a}_{\boldsymbol{y}} = \hat{b}_{\boldsymbol{y}} + \int d\boldsymbol{x} \hspace{0.5mm} F_-^\Lambda(\boldsymbol{y}-\boldsymbol{x}) \hat{b}_{\boldsymbol{x}}^\dagger,
\end{equation}
where $F_-^\Lambda(\boldsymbol{y}-\boldsymbol{x}) := \frac14 \int_{|\boldsymbol{k}|<\Lambda} \frac{d\boldsymbol{k}}{(2\pi)^n} e^{i \boldsymbol{k} \cdot (\boldsymbol{y}-\boldsymbol{x})} (\boldsymbol{k}/k_c)^2$ is one of the non-local integral kernels from \eqref{eq:yx_bbv} after the approximation.
Therefore, we see that $\hat{a}_{\boldsymbol{y}}$ and $\hat{b}_{\boldsymbol{y}}$ do not coincide in the non-relativistic regime, but differ by a term which is second order in $\Lambda/k_c$.
This second order term also shows that $\hat{a}_{\boldsymbol{y}}$ continues to act non-locally in $\boldsymbol{x}$ after the approximation.
Furthermore, introducing a cutoff aggravates this non-locality.
Whereas the original function, $F_-(\boldsymbol{y}-\boldsymbol{x}) \sim e^{-k_c |\boldsymbol{y}-\boldsymbol{x}|}$, decays exponentially beyond the Compton scale, the introduction of the cutoff causes this integral kernel to decay only polynomially, $F_-^\Lambda(\boldsymbol{y}-\boldsymbol{x}) \sim 1/|\boldsymbol{y}-\boldsymbol{x}|^{(n+1)/2}$.
Although the situation may appear asymmetric between $F_+^\Lambda$ and $F_-^\Lambda$, it turns out one can also write $F_+^\Lambda$ as a polynomially-decaying integral kernel, but it is easy to check that $F_+^\Lambda(\boldsymbol{y}-\boldsymbol{x}) := \int_{|\boldsymbol{k}|<\Lambda} \frac{d\boldsymbol{k}}{(2\pi)^n} e^{i \boldsymbol{k} \cdot (\boldsymbol{y}-\boldsymbol{x})}$ acts as the identity on bandlimited functions (i.e., it is a reproducing kernel).
Hence, in particular, $\hat{b}_{\boldsymbol{y}} = \int d\boldsymbol{x} F_+^\Lambda(\boldsymbol{y}-\boldsymbol{x}) \hat{b}_{\boldsymbol{x}}$ (keeping in mind equation \eqref{eq:bx_NR}).
This suggests that the cutoff $|\boldsymbol{k}|<\Lambda$ introduces a fundamental non-locality to both the labels $\boldsymbol{x}$ and $\boldsymbol{y}$.

This non-locality is a general feature of bandlimited functions.
A first simple observation concerning localizability is the fact that bandlimited functions cannot have compact support in position space, and hence they must exhibit some degree of non-locality.
This is a consequence of the fact that bandlimited functions are compactly supported in momentum space and of Benedicks' theorem \cite{Benedicks1985}, which limits the mutual localizability in position and momentum space (and closely related to the uncertainty principle).
Hence, we will see that the wavefunctions we obtain will exhibit an intrinsic degree of non-locality.
Further, the cutoff not only affects the operators labeled by $\boldsymbol{y}$, but also the local operators labeled by $\boldsymbol{x}$ due to the modified definition \eqref{eq:bx_NR}.

Not only do the two localization schemes differ under the approximation, but the fact that the annihilation and creation operators mix implies that they continue to generate different Fock spaces.
We can see the effects of this through the construction of so-called `quasi-local' states \cite{VazquezEtal2014} by acting the local creation operators on the global vacuum.
For example, one can write a single-particle quasi-local state as $\ket{\Psi} := \int d\boldsymbol{x} \hspace{0.5mm} \Psi(\boldsymbol{x}) \hat{b}_{\boldsymbol{x}}^\dagger \ket{0}_G$.
Notice that this agrees with the single-particle states of $\mathcal{F}[B(\Lambda)]$, since the Bogoliubov transformation \eqref{eq:yx_bbv_NR} implies $\hat{b}_{\boldsymbol{x}}^\dagger = \hat{a}_{\boldsymbol{x}}^\dagger - \int d\boldsymbol{y} F_-^\Lambda(\boldsymbol{x}-\boldsymbol{y}) \hat{a}_{\boldsymbol{y}}$, and so
\begin{equation}
  \ket{\Psi} = \int d\boldsymbol{x} \hspace{1mm} \Psi(\boldsymbol{x}) \hat{b}_{\boldsymbol{x}}^\dagger \ket{0}_G = \int d\boldsymbol{x} \hspace{1mm} \Psi(\boldsymbol{x}) \hat{a}_{\boldsymbol{x}}^\dagger \ket{0}_G,
\end{equation}
since $\hat{a}_{\boldsymbol{y}} \ket{0}_G = 0$.
Therefore, the discrepancy between the two schemes is irrelevant for single-particle states.
However, the difference will indeed manifest itself for higher particle-number states.
For example:
\begin{align}
  \int d\boldsymbol{x}_1 d\boldsymbol{x}_2 \hspace{1mm} \Psi(\boldsymbol{x}_1,\boldsymbol{x}_2) \hat{b}_{\boldsymbol{x}_1}^\dagger \hat{b}_{\boldsymbol{x}_2}^\dagger \ket{0}_G =& \int d\boldsymbol{x}_1 d\boldsymbol{x}_2 \hspace{1mm} \Psi(\boldsymbol{x}_1,\boldsymbol{x}_2) \hat{a}_{\boldsymbol{x}_1}^\dagger \hat{a}_{\boldsymbol{x}_2}^\dagger \ket{0}_G \nonumber \\
  &- \int d\boldsymbol{x}_1 d\boldsymbol{x}_2 \hspace{1mm} \Psi(\boldsymbol{x}_1,\boldsymbol{x}_2) F_-^\Lambda(\boldsymbol{x}_1-\boldsymbol{x}_2) \ket{0}_G.
\end{align}
Hence acting twice with the local creation operators generates a combination of a two-particle state and the global vacuum of $\mathcal{F}[B(\Lambda)]$.
It is generally the case that $\hat{b}_{\boldsymbol{x}_1}^\dagger \cdots \hat{b}_{\boldsymbol{x}_N}^\dagger \ket{0}_G$ will have non-zero components in particle-number sectors lower than $N$, in contrast to $\hat{a}_{\boldsymbol{x}_1}^\dagger \cdots \hat{a}_{\boldsymbol{x}_N}^\dagger \ket{0}_G$ which is an $N$-particle state in $\mathcal{F}[B(\Lambda)]$.

Hence we can conclude that the two schemes do not coincide in the non-relativistic regime, and thus we are forced to choose between them to characterize the localizability of particles and wavefunctions in the rendition of NRQM that we obtain after the approximation.
This choice is dictated by how closely each scheme can be used to recover the characteristic features of NRQM.
Of course, one of the key physical differences between NRQM and QFT is that NRQM is a theory of a fixed number of particles.
Since the number operator of the non-relativistic scheme commutes with the Hamiltonian (as opposed to that of the local operators), this clearly singles out the non-relativistic scheme as the appropriate choice.
Hence, this indeed justifies the use of the term \emph{non-relativistic} for this scheme.

We remark that this discrepancy has also implicitly appeared in other treatments of the non-relativistic approximation.
For example, in \cite{Bjorken-Drell1964}, a Foldy-Wouthuysen transformation is applied before the approximation in order to decouple the evolution of the local operators, $\hat{b}_{\boldsymbol{x}}$ and $\hat{b}_{\boldsymbol{x}}^\dagger$.
One can show that the Foldy-Wouthuysen transformation of \cite{Bjorken-Drell1964} is exactly the Bogoliubov transformation \eqref{eq:yx_bbv}, mapping the local to the non-relativistic operators.
Hence, our description of NRQM (aside from our explicit treatment of the cutoff) in terms of the non-relativistic operators is consistent with these previous works, where the transformation is applied manually in traversing from QFT to NRQM.
Although this transformation can be viewed as simply a change of representation \cite{Bjorken-Drell1964,Costella-McKellar1995}, the fact remains that the NRQM operators are non-local combinations of the field operators (and mix the local annihilation and creation operators).
In \cite{Costella-McKellar1995}, the local and non-relativistic representations (in the spin-1/2 case) were dubbed ``the two faces of the electron.''

\subsection{Salvaging NRQM from QFT}

Now we will proceed to examine which aspects of NRQM can be salvaged from the Klein-Gordon QFT after the non-relativistic approximation.
First, in addition to the non-relativistic approximation, we must restrict attention to a particular $N$-particle subspace of $\mathcal{F}[B(\Lambda)]$.
The wavefunctions of NRQM of $N$-particles will be wavepacket states of the form
\begin{equation}\label{eq:NR_wvfn}
  \ket{\Psi} := \frac{1}{\sqrt{N!}} \int d\boldsymbol{y}_1 \cdots d\boldsymbol{y}_N \hspace{1mm} \Psi(\boldsymbol{y}_1, \dots, \boldsymbol{y}_N) \hspace{1mm} \hat{a}_{\boldsymbol{y}_1}^\dagger \cdots \hat{a}_{\boldsymbol{y}_N}^\dagger \ket{0}_G,
\end{equation}
where the factor of $1/\sqrt{N!}$ is added for convenience.

Typically in such discussions, one then naturally proceeds to consider time evolution of these states.
First, let us examine the restriction of the Hamiltonian operator associated with \eqref{eq:ham} to the subspace $\mathcal{F}[B(\Lambda)]$.
Expanding the dispersion relation to second order in $\Lambda/k_c$, we have
\begin{align}
  \hat{H} \big|_{\mathcal{F}[B(\Lambda)]} &= \frac12 \int_{|\boldsymbol{k}|<\Lambda} \frac{d\boldsymbol{k}}{(2\pi)^n} \left( mc^2 + \frac{\hbar^2 \boldsymbol{k}^2}{2m} \right) ( \hat{a}_{\boldsymbol{k}}^\dagger \hat{a}_{\boldsymbol{k}} + \hat{a}_{\boldsymbol{k}} \hat{a}_{\boldsymbol{k}}^\dagger ) \\
  &= \frac12 \int d\boldsymbol{y} \left[ mc^2 ( \hat{a}_{\boldsymbol{y}}^\dagger \hat{a}_{\boldsymbol{y}} + \hat{a}_{\boldsymbol{y}} \hat{a}_{\boldsymbol{y}}^\dagger ) - \frac{\hbar^2}{2m} ( \hat{a}_{\boldsymbol{y}}^\dagger \boldsymbol{\nabla}^2 \hat{a}_{\boldsymbol{y}} + \hat{a}_{\boldsymbol{y}} \boldsymbol{\nabla}^2 \hat{a}_{\boldsymbol{y}}^\dagger ) \right].\label{eq:NR_ham}
\end{align}
We see that in fact it takes the form of a non-relativistic Hamiltonian, with an appropriate kinetic terms plus a mass energy density term.

Recall that the full relativistic Hamiltonian is local in space (labeled by $\boldsymbol{x}$),
\begin{align}
  \hat{H} &= \frac12 \int d\boldsymbol{x} \hspace{1mm} [ c^2 \hat{\Pi}(\boldsymbol{x})^2 + ( \boldsymbol{\nabla} \hat{\Phi}(\boldsymbol{x}) )^2 + k_c^2 \hat{\Phi}(\boldsymbol{x})^2 ] \\
  &= \frac12 \int d\boldsymbol{x} \left[ mc^2 ( \hat{b}_{\boldsymbol{x}}^\dagger \hat{b}_{\boldsymbol{x}} + \hat{b}_{\boldsymbol{x}} \hat{b}_{\boldsymbol{x}}^\dagger ) - \frac{\hbar^2}{2m} ( \hat{b}_{\boldsymbol{x}} + \hat{b}_{\boldsymbol{x}}^\dagger ) \boldsymbol{\nabla}^2 ( \hat{b}_{\boldsymbol{x}} + \hat{b}_{\boldsymbol{x}}^\dagger ) \right].
\end{align}
However, the full relativistic Hamiltonian written in terms of the non-relativistic localization scheme is
\begin{equation}\label{eq:ay_ham}
  \hat{H} = \frac12 \int \frac{d\boldsymbol{k}}{(2\pi)^n} \hbar \omega_{\boldsymbol{k}} ( \hat{a}_{\boldsymbol{k}}^\dagger \hat{a}_{\boldsymbol{k}} + \hat{a}_{\boldsymbol{k}} \hat{a}_{\boldsymbol{k}}^\dagger ) = \frac12 \int d\boldsymbol{y}' d\boldsymbol{y} f(\boldsymbol{y}'-\boldsymbol{y}) ( \hat{a}_{\boldsymbol{y}'}^\dagger \hat{a}_{\boldsymbol{y}} + \hat{a}_{\boldsymbol{y}} \hat{a}_{\boldsymbol{y}'}^\dagger ),
\end{equation}
where $f(\boldsymbol{y}'-\boldsymbol{y}) := \int \frac{d\boldsymbol{k}}{(2\pi)^n} \hbar \omega_{\boldsymbol{k}} e^{i \boldsymbol{k} \cdot ( \boldsymbol{y}' - \boldsymbol{y} ) }$ is a non-local integral kernel.
Comparing \eqref{eq:NR_ham} with \eqref{eq:ay_ham}, we see that the non-relativistic approximation seems to remove the non-locality associated with the integral kernel $f$, and we obtain a Hamiltonian which appears local in $\boldsymbol{y}$ space.
However, this observation is somewhat deceptive, since the cutoff $|\boldsymbol{k}|<\Lambda$ introduces non-locality in the functions over $\boldsymbol{y}$.

Since the $N$-particle subspace of $\mathcal{F}[B(\Lambda)]$ is preserved in time, the time-dependence of the wavepacket states \eqref{eq:NR_wvfn} can be absorbed into the smearing function which encodes the coefficients of the state in this subspace, i.e.,
\begin{equation}
  \ket{\Psi(t)} := \frac{1}{\sqrt{N!}} \int d\boldsymbol{y}_1 \cdots d\boldsymbol{y}_N \hspace{1mm} \Psi(\boldsymbol{y}_1, \dots, \boldsymbol{y}_N; t) \hspace{1mm} \hat{a}_{\boldsymbol{y}_1}^\dagger \cdots \hat{a}_{\boldsymbol{y}_N}^\dagger \ket{0}_G.
\end{equation}
One can then translate the abstract time evolution of this state under the Hamiltonian \eqref{eq:NR_ham}, $i \hbar \frac{d}{dt} \ket{\Psi(t)} = \hat{H} \ket{\Psi(t)}$, into a differential equation for the smearing function $\Psi(\boldsymbol{y}_1, \dots, \boldsymbol{y}_N; t)$.
It is easy to see that the smearing function indeed satisfies the quantum mechanical Schr\"odinger equation,
\begin{equation}
  i \hbar \partial_t \Psi(\boldsymbol{y}_1, \dots, \boldsymbol{y}_N; t) = \left( E_0 + mc^2 N - \frac{\hbar^2}{2m} \sum_{i=1}^N \boldsymbol{\nabla}_i^2 \right) \Psi(\boldsymbol{y}_1, \dots, \boldsymbol{y}_N; t),
\end{equation}
where $E_0$ is the zero-point energy.
Note that we make contact between QFT and NRQM by associating the smearing function in the $N$-particle subspace with the non-relativistic wavefunction, and not the field operator as would be suggested by the archaic use of the term `second quantization'.
Indeed, this is the only manner in which one could arrive at a multi-particle Schr\"odinger equation.

However, NRQM exhibits more structure than simply the Schr\"odinger equation; we need to further justify how the smearing function imitates the familiar NRQM wavefunction under this approximation.
For instance, the wavefunction $\Psi(\boldsymbol{y}_1, \dots, \boldsymbol{y}_N)$ would typically represent the probability amplitude for measuring $N$ particles at the points $\boldsymbol{y}_1, \dots, \boldsymbol{y}_N$.
Do the wavefunctions we obtain here have such an interpretation?
Does the label $\boldsymbol{y}_i$ correspond to an eigenvalue of a position-type operator satisfying the Heisenberg algebra?
Do we recover a Born rule and state-update rule for these wavefunctions?

Ordinarily a quantum mechanical wavefunction is defined over space by $\Psi(\boldsymbol{x}_1, \dots, \boldsymbol{x}_N) := \langle \boldsymbol{x}_1, \dots, \boldsymbol{x}_N | \Psi \rangle$, where $\ket{\boldsymbol{x}_1, \dots, \boldsymbol{x}_N}$ is an element of a position basis provided by the position operators, $\boldsymbol{\hat{x}}_1, \dots, \boldsymbol{\hat{x}}_N$, of the $N$ particles.
In order to describe bosonic (or fermionic) particles, this wavefunction must be symmetrized (or antisymmetrized) over the particle labels.
In QFT, we do not naturally have position operators to provide us with such a basis; the arguments, $\boldsymbol{y}_i$, of the smearing function rather label the infinite degrees of freedom of the non-relativistic localization scheme.
For the wavepackets \eqref{eq:NR_wvfn} in the bosonic Klein-Gordon field, the symmetrization of the particle labels occurs automatically due to $[ \hat{a}_{\boldsymbol{y}}^\dagger, \hat{a}_{\boldsymbol{y}'}^\dagger ] = 0$.
However, after this symmetrization, in both the QFT and NRQM cases, one cannot have a physical position operator $\boldsymbol{\hat{x}}_i$ for the $i^\text{th}$ particle as it would not map the symmetric subspace into itself.
Despite this, on physical grounds there should be a means of describing measurements of distance between identical particles.
Although this is clearly an important issue, we will not resolve it here, since the problem is also present in NRQM and not a distinct feature of localizability issues in QFT.
At least, we can define a center-of-mass position operator which is conjugate to the total momentum operator of the QFT, namely
\begin{equation}
  \boldsymbol{\hat{X}} := \hat{N}^+ \int d\boldsymbol{y} \hspace{1mm} \boldsymbol{y} \hspace{1mm} \hat{a}_{\boldsymbol{y}}^\dagger \hat{a}_{\boldsymbol{y}},
\end{equation}
where $\hat{N}^+$ is the pseudo-inverse of $\hat{N}$.
The formal eigenstates of this position operator are:
\begin{equation}\label{eq:posn_eig}
  \ket{\boldsymbol{y}_1, \dots, \boldsymbol{y}_N} \equiv \frac{1}{\sqrt{N!}} \hspace{1mm} \hat{a}_{\boldsymbol{y}_1}^\dagger \cdots \hat{a}_{\boldsymbol{y}_N}^\dagger \ket{0}_G,
\end{equation}
with eigenvalues $\frac1N \sum_{i=1}^N \boldsymbol{y}_i$.
This operator satisfies the Heisenberg algebra with the total momentum operator:
\begin{align}
  \left[ \boldsymbol{\hat{X}}, \boldsymbol{\hat{P}}^T \right] &= \left[ \hat{N}^+ \int d\boldsymbol{y} \hspace{1mm} \boldsymbol{y} \hspace{1mm} \hat{a}_{\boldsymbol{y}}^\dagger \hat{a}_{\boldsymbol{y}} , \int_{|\boldsymbol{k}|<\Lambda} \frac{d\boldsymbol{k}}{(2\pi)^n} \hbar \boldsymbol{k}^T \hat{a}_{\boldsymbol{k}}^\dagger \hat{a}_{\boldsymbol{k}} \right] \nonumber \\
  &= i \hbar \boldsymbol{\mathds{1}}_n ( \mathds{1}_{\mathcal{F}[B(\Lambda)]} - \ket{0}_G \bra{0}_G ),
\end{align}
where $\boldsymbol{\mathds{1}}_n$ is used to denote the fact that the Heisenberg algebra is satisfied component-wise for each of the $n$ spatial components of these operators, and the projector $( \mathds{1}_{\mathcal{F}[B(\Lambda)]} - \ket{0}_G \bra{0}_G )$ represents the identity on the Fock space, except the zero-particle subspace which is annihilated by both $\boldsymbol{\hat{X}}$ and $\boldsymbol{\hat{P}}$.

Alas, at present it is not clear how to concoct a complete set of position observables for the non-relativistic wavefunctions we obtain from the QFT (except the single-particle subspace where the above center-of-mass position operator suffices).
Nevertheless, we can build sets of measurement operators associated with asking questions of the kind: ``What is the probability of finding one of the $N$ particles in a region $\Delta \subset \mathds{R}^n$?''
Such measurements can serve to approximately characterize the localizability of these states (not fully, since there remains a discrepancy between $\boldsymbol{x}$-space and $\boldsymbol{y}$-space).
The measurement operators can be constructed in the $N$-particle subspace using the formal position eigenvectors \eqref{eq:posn_eig}:
\begin{equation}\label{eq:posn_POVM}
  P_\Delta^{(N)} := \int_\Delta d\boldsymbol{y}_1 \int d\boldsymbol{y}_2 \cdots d\boldsymbol{y}_N \hspace{1mm} \ket{\boldsymbol{y}_1, \dots, \boldsymbol{y}_N} \bra{\boldsymbol{y}_1, \dots, \boldsymbol{y}_N}.
\end{equation}
The expectation value of these operators over an $N$-particle state \eqref{eq:NR_wvfn} yields
\begin{equation}
  \bra{\Psi} P_\Delta^{(N)} \ket{\Psi} = \int_\Delta d\boldsymbol{y}_1 \int d\boldsymbol{y}_2 \cdots d\boldsymbol{y}_N \hspace{1mm} | \Psi(\boldsymbol{y}_1, \dots, \boldsymbol{y}_N) |^2.
\end{equation}
If one is agnostic as to the total number of particles, then one could use the operator $P_\Delta = \sum_{N=1}^\infty P_\Delta^{(N)}$ defined over the total bandlimited Fock space.
It is straightforward to construct measurement operators for similar types of questions.

Hence we see that these wavefunctions are indeed associated with probability amplitudes associated with these kind of measurements which aim to localize these particles.
However, interpreting these measurements is not as straightforward as it may seem.
As we mentioned above, the cutoff causes the wavefunctions to exhibit a kind of non-locality, and this gives these measurements some unintuitive features.
First, note that these measurement operators are not projectors, as one can easily verify $( P_\Delta^{(N)} )^2 \neq P_\Delta^{(N)}$.
This is a consequence of the fact that the formal eigenvectors \eqref{eq:posn_eig} of the position operator are not orthogonal because of the cutoff.
For example,
\begin{equation}
  \langle \boldsymbol{y}' | \boldsymbol{y} \rangle = \int_{|\boldsymbol{k}|<\Lambda} \frac{d\boldsymbol{k}}{(2\pi)^n} e^{i \boldsymbol{k} \cdot (\boldsymbol{y}'-\boldsymbol{y})} \neq \delta(\boldsymbol{y}'-\boldsymbol{y}).
\end{equation}
Hence this collection of formal position eigenvectors form a frame rather than an orthonormal basis.
Note that the possibility that these formal eigenvectors do not form an orthonormal basis is due to the fact that the position operator we constructed is not essentially self-adjoint\footnote{One can see this by observing that the bandlimitation causes $\boldsymbol{\hat{X}}$ to have a finite minimum uncertainty, hence it cannot have any eigenvectors \cite{Kempf1994a,Kempf-Mangano-Mann1995}. For example, for a single-particle in one dimension, the bandlimitation causes a maximum uncertainty $\Delta \hat{P} \leq 2\Lambda$, thus the uncertainty principle forces $\Delta \hat{X} \geq \hbar/4\Lambda$.} \cite{Kempf1994a,Kempf-Mangano-Mann1995}.
Nevertheless, the operators $P_\Delta^{(N)}$ are clearly positive, hence the measurements we have described are associated with a POVM-type measurement.

Due to the fact that these measurement operators are elements of a POVM, and not projectors, prevents us from being able to write a state-update rule because we do not have the Kraus operators associated with the measurement.
Generally, the measurement theory which corresponds to ``observables'' of this kind is unclear.
Note that such objects do not only appear in such esoteric contexts; indeed, one also arrives in a similar situation when dealing with momentum and Hamiltonian operators for a particle in an infinite potential well (however, in some cases there are physically-motivated ways to resolve the difficulties \cite{Bonneau-Faraut-Valent2001,Fewster1993}).
We leave further investigation into these issues as future work. 

In a relativistic setting, a theorem of Hegerfeldt \cite{Hegerfeldt1998a,Hegerfeldt1998b,Hegerfeldt1974} trivializes POVM elements that one might wish to associate with spatial regions, similar to how Malament's theorem \cite{Malament1996} trivializes projectors.
The key relativistic assumption is the requirement that there should be a finite speed of propagation \cite{Hegerfeldt1998a,Hegerfeldt1998b,Hegerfeldt1974}.
This is violated in both cases by energy positivity \cite{Hegerfeldt1998a,Hegerfeldt1998b,Hegerfeldt1974,Halvorson2001}.
The fact that we can construct non-trivial POVM elements \eqref{eq:posn_POVM} in the non-relativistic regime suggests that this requirement from relativity contributes to trivializing the spatial POVMs.
Of course, the localizability expressed in terms of these elements will suffer from superluminal propagation.
In our case, technically this is due to the fact that the ultraviolet cutoff breaks the Lorentz-invariance of the theory.
However, this is not an issue since superluminal propagation is also an aspect of NRQM.
The point is that we recover tools for characterizing localizability of states, despite their `pathological' propagation which can be attributed to the non-relativistic approximation.

\subsection{Local degrees of freedom and sampling theory}\label{subsec:sampling}

An aspect of the recovered theory which is incongruous with the usual structure of NRQM is that the space consists of wavefunctions which are bandlimited, not the full space $L_2(\mathds{R}^n)$.
Of course this makes sense because physically we know NRQM does not apply at large velocities, so it should disagree with QFT in this regime.
However, localizability of bandlimited functions is a delicate issue; this property has important implications regarding the arrangement of the spatial degrees of freedom in the field theory.
These implications will be particularly important for the discussion in Section~\ref{sec:entanglement}.
In this subsection, we will describe a framework that can be used to interpret these features.

Initially we had labeled points in space by $\boldsymbol{x}$, and the corresponding degrees of freedom through $\hat{\Phi}(\boldsymbol{x})$ and $\hat{\Pi}(\boldsymbol{x})$.
We then constructed the local Fock space through taking a tensor product over $\boldsymbol{x}$ labeling these degrees of freedom.
The bandlimited Fock space we used for the non-relativistic approximation has a tensor product structure over $\boldsymbol{k}$, i.e., $\mathcal{F}[B(\Lambda)] \cong \otimes_{|\boldsymbol{k}|<\Lambda} L_2(\mathds{C},d\Phi_{\boldsymbol{k}})$.
Is there an analogue of the $\boldsymbol{x}$ tensor product decomposition (even if it is a unitarily-inequivalent representation)?
As we shall see, it is not simply a space of the form $\otimes_{\boldsymbol{x}} \mathcal{H}_{\boldsymbol{x}}$.
This can be deduced from the fact that the operators $\hat{\Phi}(\boldsymbol{x})$ and $\hat{\Pi}(\boldsymbol{x})$ act non-locally after introducing the cutoff, which can be seen from the new commutation relations:
\begin{equation}
  [ \hat{\Phi}(\boldsymbol{x}), \hat{\Pi}(\boldsymbol{x}') ] = i \hbar \int_{|\boldsymbol{k}|<\Lambda} \frac{d\boldsymbol{k}}{(2\pi)^n} e^{i \boldsymbol{k} \cdot (\boldsymbol{x}-\boldsymbol{x}')}.
\end{equation}
For example, for $n=1$, this becomes $[ \hat{\Phi}(x), \hat{\Pi}(x') ] = i \hbar \frac{\Lambda}{\pi} \sinc [ \Lambda (x-x') ]$, which decays $\sim 1/|x-x'|$.
Hence, the operators cannot act on a single factor of a Hilbert space of the form $\otimes_{\boldsymbol{x}} \mathcal{H}_{\boldsymbol{x}}$.
In light of this, is it possible to find an analogue characterization for the spatial degrees of freedom of the field theory?

For the purpose of clarity in describing the main features of sampling theory, in this subsection we will employ a cutoff on each of the coordinates of the wavevector $\boldsymbol{k}$, rather than the spherically-symmetric cutoff $|\boldsymbol{k}|<\Lambda$.
We will denote this cutoff by $\| \boldsymbol{k} \|_\infty < \Lambda$.
Clearly this will not affect the general features of the expansions under the non-relativistic approximation, as these norms are equivalent ($\| \boldsymbol{k} \|_\infty \leq \| \boldsymbol{k} \|_2 \leq \sqrt{n} \| \boldsymbol{k} \|_\infty$, where $|\boldsymbol{k}| \equiv \| \boldsymbol{k} \|_2$), and the only condition we required is $\Lambda \ll k_c$.
It is still possible to prove statements regarding the sampling theory for the spherically-symmetric case \cite{Landau1967,Jerri1977}, but cutting off each coordinate reduces the problem to a product of one-dimensional cases for which one can write down explicit formulas which will be pedagogically useful and will suffice for the present exposition.

The question of characterizing the spatial degrees of freedom of bandlimited fields was answered by Shannon sampling theory \cite{Shannon1948,Shannon1949,Nyquist1928} for classical bandlimited functions, and can be extended to quantum field theory \cite{Kempf1994b,Kempf1997,Kempf2000,Pye-Donnelly-Kempf2015}.
(\cite{Pye-Donnelly-Kempf2015} also contains a discussion about localizability in bandlimited quantum fields, albeit the context is a model for Planck-scale physics.)
The basic result of sampling theory for a function, $f$, with a cutoff $\| \boldsymbol{k} \|_\infty < \Lambda$, can be summarized by the following reconstruction formula:
\begin{equation}\label{eq:sampling}
  f(\boldsymbol{x}) = \sum_{\boldsymbol{m} \in \mathds{Z}^n} K(\boldsymbol{x},\boldsymbol{x}_{\boldsymbol{m}}^{(\boldsymbol{\alpha})}) f(\boldsymbol{x}_{\boldsymbol{m}}^{(\boldsymbol{\alpha})}), \quad \text{where} \quad K(\boldsymbol{x},\boldsymbol{x}_{\boldsymbol{m}}^{(\boldsymbol{\alpha})}) := \prod_{i=1}^n \sinc [ \Lambda ( x^i - (x^i)_{n^i}^{(\alpha^i)}  ) ]
\end{equation}
where $\boldsymbol{\alpha} \in [0,1)^n$ is an arbitrary parameter, $\{ \boldsymbol{x}_{\boldsymbol{m}}^{(\boldsymbol{\alpha})} := \pi (\boldsymbol{m}-\boldsymbol{\alpha})/\Lambda \}_{\boldsymbol{m} \in \mathds{Z}^n}$ is a set of \emph{sampling points} (or a \emph{sampling lattice}), and we use the notation $x^i$ to denote the $i^{th}$ component of the vector $\boldsymbol{x}$.
The essence of this formula is that the value of the function at \emph{any} $\boldsymbol{x} \in \mathds{R}^n$ is completely determined by the values of the function on a sampling lattice.
The values of a bandlimited function on a sampling lattice embody the independent spatial degrees of freedom.
Any one of the $\boldsymbol{\alpha}$-parametrized family of sampling lattices will serve for the above reconstruction formula, hence one can choose the values on any of the lattices as the degrees of freedom.
One can think of the above reconstruction formula as stating that the sampling kernels, $K(\boldsymbol{x},\boldsymbol{x}_{\boldsymbol{m}}^{(\boldsymbol{\alpha})})$, centered at the sample points of one of these lattices form an orthonormal basis for the space of bandlimited functions, and that the coefficients $f(\boldsymbol{x}_{\boldsymbol{m}}^{(\boldsymbol{\alpha})})$ specify a particular function in this space.
The arbitrary parameter $\boldsymbol{\alpha}$ indicates that there is a parametrized family of bases which are simply translated versions of one another.

It is straightforward to show that the above reconstruction formula extends to bandlimited quantum fields \cite{Pye-Donnelly-Kempf2015}: $\hat{\Phi}(\boldsymbol{x}) = \sum_{\boldsymbol{m} \in \mathds{Z}^n} K(\boldsymbol{x},\boldsymbol{x}_{\boldsymbol{m}}^{(\boldsymbol{\alpha})}) \hat{\Phi}(\boldsymbol{x}_{\boldsymbol{m}}^{(\boldsymbol{\alpha})})$.
Supporting the claim that the function values on the sampling lattice exhibit the independent spatial degrees of freedom of the field theory, we note that these operators satisfy canonical commutation relations (up to a multiplicative factor) between points on the same sampling lattice (i.e., same $\boldsymbol{\alpha}$): $[ \hat{\Phi}(\boldsymbol{x}_{\boldsymbol{m}}^{(\boldsymbol{\alpha})}), \hat{\Pi}(\boldsymbol{x}_{\boldsymbol{m}'}^{(\boldsymbol{\alpha})}) ] = i \hbar \left( \frac{\Lambda}{\pi} \right)^n \delta_{\boldsymbol{m} \boldsymbol{m}'}$.
Hence, the analogue of the local Hilbert space for the spatial degrees of freedom in this bandlimited theory is: $\mathcal{F}_L^\Lambda := \otimes_{\boldsymbol{m} \in \mathds{Z}^n} L_2(\mathds{R},d\Phi(\boldsymbol{x}_{\boldsymbol{m}}^{(\boldsymbol{\alpha})})) \cong \mathcal{F}[\ell_2^n]$, i.e., the local tensor product structure is over lattice points.
We also not that one obtains a different tensor product structure for each of the sampling lattices.
However, one can easily use the reconstruction formula to show that the Bogoliubov transformation between different lattices contains no mixing between the annihilation and creation operators:
\begin{equation}
  \hat{b}_{\boldsymbol{x}_{\boldsymbol{m}}^{(\boldsymbol{\alpha})}} = \sum_{\boldsymbol{m}' \in \mathds{Z}^n} K(\boldsymbol{x}_{\boldsymbol{m}}^{(\boldsymbol{\alpha})}, \boldsymbol{x}_{\boldsymbol{m}'}^{(\boldsymbol{\alpha}')}) \hat{b}_{\boldsymbol{x}_{\boldsymbol{m}'}^{(\boldsymbol{\alpha}')}},
\end{equation}
where $\hat{b}_{\boldsymbol{x}_{\boldsymbol{m}}^{(\boldsymbol{\alpha})}} := \sqrt{\frac{m}{2\hbar^2}} \hat{\Phi}(\boldsymbol{x}_{\boldsymbol{m}}^{(\boldsymbol{\alpha})}) + \frac{i}{\sqrt{2m}} \hat{\Pi}(\boldsymbol{x}_{\boldsymbol{m}}^{(\boldsymbol{\alpha})})$.
Hence regardless of the lattice which is chosen, one arrives at the same Fock space.

How does this `local' Fock space associated with the sample points compare to the bandlimited Fock space, $\mathcal{F}[B(\Lambda)]$, in which we constructed the $N$-particle spaces of NRQM?
We have already demonstrated in Subsection~\ref{subsec:two_schemes_NR} that they are not the same.
In Appendix~\ref{apdx:bbv}, we show that they remain unitarily inequivalent (due to an infrared divergence), as in the case of the full local and global Fock spaces.

We also note that the non-relativistic operators also exhibit the bandlimitation property.
For example, one can similarly write
\begin{equation}
  \hat{a}_{\boldsymbol{y}} = \sum_{\boldsymbol{p} \in \mathds{Z}^n} K(\boldsymbol{y},\boldsymbol{y}_{\boldsymbol{p}}^{(\boldsymbol{\beta})}) \hat{a}_{\boldsymbol{y}_{\boldsymbol{p}}^{(\boldsymbol{\beta})}}.
\end{equation}
Of course, because of the remaining discrepancy between the $\boldsymbol{x}$ and $\boldsymbol{y}$ labels, the sampling lattices in these two spaces are different.
That is, one can show the non-local kernel $F_-^\Lambda$ in \eqref{eq:yx_bbv_NR} is such that $F_-^\Lambda(\boldsymbol{y}_{\boldsymbol{p}}^{(\boldsymbol{\beta})}-\boldsymbol{x}_{\boldsymbol{m}}^{(\boldsymbol{\alpha})}) \not\sim \delta_{\boldsymbol{p},\boldsymbol{m}}$ for any $\boldsymbol{y}$-sample point $\boldsymbol{y}_{\boldsymbol{p}}^{(\boldsymbol{\beta})}$ and any $\boldsymbol{x}$-sample point $\boldsymbol{x}_{\boldsymbol{m}}^{(\boldsymbol{\alpha})}$.

Therefore, we see that after the bandlimitation introduced in order to enact the non-relativistic approximation, the resulting tensor product structures akin to the original $\boldsymbol{x}$ and $\boldsymbol{y}$ tensor product structures are those taken over lattice points.
The degrees of freedom of the field theory after this approximation can be identified with the field values at the sample points in $\boldsymbol{x}$-space.
Similarly, the wavefunctions of the NRQM we obtain in $\boldsymbol{y}$-space also exhibit the sampling property in each coordinate, hence these wavefunctions are determined by their values on a sampling lattice.
Thus, the representation of the wavefunction over all of $\boldsymbol{y}$-space is a redundant description.
This redundancy is related to the fact that the formal position eigenvectors, $\ket{\boldsymbol{y}_1, \dots, \boldsymbol{y}_N}$, form a frame rather than an orthonormal basis.
In fact, this collection of states can be seen as a union of orthonormal bases associated with the independent points of the sampling lattices \cite{Pye-Donnelly-Kempf2015}.

\section{Is the Unruh effect relativistic?}\label{sec:entanglement}

In the previous sections, we showed that one could define bandlimited versions of the local annihilation and creation operators acting on the bandlimited subspace of the global Fock space.
Although the corresponding number operator does not provide a counting of the particles associated with this Fock space, one can still ask about the fate of the Reeh-Schlieder theorem under the non-relativistic approximation.
We do not aim here to construct full analogues of the assumptions which lead to the Reeh-Schlieder theorem in the non-relativistic regime, but rather use the relation with vacuum entanglement \cite{Redhead1995} and investigate this instead.

One could also ask the question, out of independent interest, whether vacuum entanglement is a feature of the relativistic regime of the field theory.
\updated{As we said in the introduction, we are not implying that ground state entanglement only occurs in relativistic systems, but rather we are asking whether it occurs only in the relativistic regime of a relativistic QFT.}
Intuitively, ground state entanglement is caused by coupling between degrees of freedom in the Hamiltonian of a theory.
However, not all couplings lead to entanglement; indeed, we saw above that the non-relativistic field operators do not exhibit ground state entanglement, yet they are coupled through derivative terms in the Hamiltonian.
Ground state entanglement is exhibited by systems which are \emph{frustrated}, i.e., whose free and interaction terms\footnote{In our case, by \emph{interaction} term we mean the term of the Hamiltonian which couples the local oscillators, i.e., the spatial derivative term.} of the Hamiltonian do not commute.
In some instances, one can concretely relate measures of frustration with bounds on ground state entanglement \cite{Dawson-Nielsen2004}.
It is known that the local degrees of freedom of a Klein-Gordon field are entangled in the ground state, but it is not intuitively obvious whether the frustration of the couplings is a relativistic effect.

Ground state entanglement in quantum field theory is neatly displayed by the Unruh temperature, $k_B T_U = \hbar a / 2\pi c$.
There are two distinct ways of thinking about the physics behind this temperature: one in terms of the stimulation of a process involving a counter-rotating wave type interaction term between an accelerated detector and the field \cite{Unruh1976} (which can be extended to more general trajectories), and the other in terms of the thermality of the reduced state on a half-space caused by entanglement \cite{Wald1994,Sorkin1983,Bombelli-Koul-Lee-Sorkin1986} (which can be extended to reduced states on more general local subsystems).
Here we are interested in the latter.
A naive inspection of the Unruh temperature suggests it should vanish in the limit $c \to \infty$, hence one would be tempted to conclude that the effect is relativistic.
(Note this is similar to the manner in which one describes the Hawking temperature, $k_B T_H = \hbar c^3 / 8\pi GM$, as exhibiting features of quantum theory, relativity, and gravity, due to the presence of $\hbar$, $c$, and $G$.)
However, as we discussed previously, the non-relativistic regime does not correspond to a limit, but rather an approximation to second order in $\Lambda/k_c$, after introducing a cutoff $\Lambda$.
Here we will examine more carefully the requirement of relativity for the presence of the Unruh effect in \updated{relativistic} quantum field theory.

\updated{
An important ingredient in the standard derivation of the Unruh effect (viewed as arising from entanglement) is the presence of a horizon for a uniformly accelerated observer, which naturally splits the field into two (entangled) subsystems.
Of course, without relativity there can be no horizons present.
However, if there were some other circumstance (within the non-relativistic regime) in which an observer would aptly trace out some local subsystem, then such an observer would perceive an Unruh-like effect if the local subsystems remain entangled in this regime.
For example, in the relativistic setting, this occurs in situations where an observer couples locally to the field for a finite period of time.
Indeed, there are many works which refer to such effects as the ``Unruh effect'' \cite{martinetti2003diamond,crispino2008unruh}.
It is this generalized sense in which we are using the term throughout the text.
Primarily this literature concerns itself with the responses of detectors coupled to the field.
Here we are investigating, in the non-relativistic regime, whether a contribution to the response of such a detector would arise from the thermality of local subsystems in the field due to the fact that they are entangled.
As we are interested in isolating the contribution of the response due to entanglement, we will simply determine whether the local subsystems of the field are entangled in this regime.
}

\subsection{Local degrees of freedom remain entangled}

Here we are identifying the Unruh effect with entanglement between local degrees of freedom.
\updated{Here we are identifying the points on a sampling lattice as the local degrees of freedom after the ``coarse-graining'' induced by the cutoff for the non-relativistic limit.
Therefore, in this context, the Unruh effect is related to entanglement between the field at these sample points.
(See also another perspective in \cite{cacciatori2009renormalized}, where the herein identified non-relativistic degrees of freedom were identified as the appropriate coarse-grained degrees of freedom for the purposes of thermodynamics.
There, the entropy of a region of space, as identified by the non-relativistic localization scheme, was calculated for a global thermal state.)
}

First, we will demonstrate that the sample points are indeed entangled in the non-relativistic regime.
We begin with the Bogoliubov transformation,
\begin{equation}
  \hat{a}_{\boldsymbol{k}} = \left( \frac{\pi}{\Lambda} \right)^{n/2} \sum_{\boldsymbol{m} \in \mathds{Z}^n} e^{-i \boldsymbol{k} \cdot \boldsymbol{x}_{\boldsymbol{m}}^{(\boldsymbol{\alpha})}} \left[ \hat{b}_{\boldsymbol{m}} + \frac14 (\boldsymbol{k}/k_c)^2 \hat{b}_{\boldsymbol{m}}^\dagger \right],
\end{equation}
where for convenience we will fix $\boldsymbol{\alpha}$ and write $\hat{b}_{\boldsymbol{m}} \equiv \left( \frac{\pi}{\Lambda} \right)^{n/2} \hat{b}_{\boldsymbol{x}_{\boldsymbol{m}}^{(\boldsymbol{\alpha})}}$.
The factor $\left( \frac{\pi}{\Lambda} \right)^{n/2}$ is included in the definition of $\hat{b}_{\boldsymbol{m}}$ so that the commutation relations are properly normalized: $[ \hat{b}_{\boldsymbol{m}}, \hat{b}_{\boldsymbol{m}'}^\dagger ] = \delta_{\boldsymbol{m} \boldsymbol{m}'}$.
The above expression can be obtained by inverting the Bogoliubov transformation \eqref{eq:bx_NR} and using the sampling formula \eqref{eq:sampling}.
Generally, given a Bogoliubov transformation of the form,
\begin{equation}
  \hat{a}_i = \sum_j ( \alpha_{ij} \hat{b}_j + \beta_{ij} \hat{b}_j^\dagger ),
\end{equation}
we can relate the two vacua by a unitary,
\begin{equation}
  \ket{0}_a = N e^{-\frac12 \sum_{j,k} (\alpha^{-1} \beta)_{jk} \hat{b}_j^\dagger \hat{b}_k^\dagger} \ket{0}_b,
\end{equation}
provided the two spaces are unitarily equivalent, and the operator $(\alpha^{-1} \beta)$ exists and is symmetric.
(One can easily show that the latter two conditions always holds for a valid Bogoliubov transformation.)
The factor $N$ is simply a normalization constant.

In our case, we have
\begin{equation}
  \alpha_{\boldsymbol{k} \boldsymbol{m}} = \left( \frac{\pi}{\Lambda} \right)^{n/2} e^{-i \boldsymbol{k} \cdot \boldsymbol{x}_{\boldsymbol{m}}^{(\boldsymbol{\alpha})}}, 
  \qquad \beta_{\boldsymbol{k} \boldsymbol{m}} = \left( \frac{\pi}{\Lambda} \right)^{n/2} e^{-i \boldsymbol{k} \cdot \boldsymbol{x}_{\boldsymbol{m}}^{(\boldsymbol{\alpha})}} \frac14 (\boldsymbol{k}/k_c)^2,
\end{equation}
and
\begin{equation}
  (\alpha^{-1} \beta)_{\boldsymbol{m} \boldsymbol{m}'} = \frac14 \int_{\| \boldsymbol{k} \|_\infty < \Lambda} \frac{d\boldsymbol{k}}{(2\Lambda)^n} e^{i \boldsymbol{k} \cdot (\boldsymbol{x}_{\boldsymbol{m}}^{(\boldsymbol{\alpha})} - \boldsymbol{x}_{\boldsymbol{m}'}^{(\boldsymbol{\alpha})})} (\boldsymbol{k}/k_c)^2,
\end{equation}
which is indeed symmetric.
However, as we mentioned above, the bandlimited local and global Fock space representations are not unitarily equivalent.
Nevertheless, we will proceed with the following formal manipulations to suggest that the local degrees of freedom remain entangled, and proceed to investigate more carefully in Subsections~\ref{subsec:osc_temp} and \ref{subsec:osc_logneg}.
We will denote the global vacuum, $\ket{0}_G^\Lambda := \otimes_{\| \boldsymbol{k} \|_\infty < \Lambda} \ket{0}_{\boldsymbol{k}}$, and the local sampling vacuum, $\ket{0}_L^\Lambda := \otimes_{\boldsymbol{m} \in \mathds{Z}^n} \ket{0}_{\boldsymbol{m}}$.
Therefore, to second order in $\| \boldsymbol{k} \|_\infty / k_c$, we can write
\begin{align}
  \ket{0}_G^\Lambda &= N \exp \left[ -\frac12 \sum_{\boldsymbol{m}, \boldsymbol{m}' \in \mathds{Z}^n} (\alpha^{-1} \beta)_{\boldsymbol{m} \boldsymbol{m}'} \hat{b}_{\boldsymbol{m}}^\dagger \hat{b}_{\boldsymbol{m}'}^\dagger \right] \ket{0}_L^\Lambda \nonumber \\
  &\sim \ket{0}_L^\Lambda - \frac12 \sum_{\boldsymbol{m}, \boldsymbol{m}' \in \mathds{Z}^n} (\alpha^{-1} \beta)_{\boldsymbol{m} \boldsymbol{m}'} \hat{b}_{\boldsymbol{m}}^\dagger \hat{b}_{\boldsymbol{m}'}^\dagger \ket{0}_L^\Lambda.
\end{align}
Note that to second order, this state is a combination the vacuum and two-particle states in the local Fock space, with the terms at higher order lying in higher particle number subspaces.
If we write the states explicitly in a tensor decomposition over lattice points, we get
\begin{equation}
  \ket{0}_G^\Lambda = \ket{0 0 \cdots} - \frac12 \sum_{\boldsymbol{m} \neq \boldsymbol{m}'} (\alpha^{-1} \beta)_{\boldsymbol{m} \boldsymbol{m}'} \ket{0 \cdots 0 1_{\boldsymbol{m}} 0 \cdots 0 1_{\boldsymbol{m}'} 0 \cdots} - \frac{1}{\sqrt{2}} \sum_{\boldsymbol{m}} (\alpha^{-1} \beta)_{\boldsymbol{m} \boldsymbol{m}} \ket{0 \cdots 0 2_{\boldsymbol{m}} 0 \cdots}.
\end{equation}

Such a state is always entangled if an off-diagonal entry of $(\alpha^{-1} \beta)$ is non-zero (as it is in our case, which can be easily verified).
This is because if one tried to write it as a state of the form,
\begin{equation}
  \ket{\psi} := \otimes_{\boldsymbol{m}} ( A_{\boldsymbol{m}} \ket{0}_{\boldsymbol{m}} + B_{\boldsymbol{m}} \ket{1}_{\boldsymbol{m}} + C_{\boldsymbol{m}} \ket{2}_{\boldsymbol{m}} ),
\end{equation}
then one arrives at a contradiction in attempting to match coefficients.
The coefficient of $\ket{0 0 \cdots}$ in $\ket{\psi}$ is $\prod_{\boldsymbol{m}} A_{\boldsymbol{m}}$ which must be 1 to second order, which means that (up to global phases) we must have $A_{\boldsymbol{m}} = 1$ for all $\boldsymbol{m}$.
Next we have the single-particle components of $\ket{\psi}$, which have coefficients $B_{\boldsymbol{m}}$ for $\ket{0 \cdots 0 1_{\boldsymbol{m}} 0 \cdots}$ and must all vanish.
Hence we require $B_{\boldsymbol{m}} = 0$ for all $\boldsymbol{m}$.
Now we have a contradiction, since if a particular off-diagonal component $(\alpha^{-1} \beta)_{\boldsymbol{m} \boldsymbol{m}'} \neq 0$ (with $\boldsymbol{m} \neq \boldsymbol{m}'$), then we require the coefficient of $\ket{0 \cdots 0 1_{\boldsymbol{m}} 0 \cdots 0 1_{\boldsymbol{m}'} 0 \cdots}$ to be both $B_{\boldsymbol{m}} B_{\boldsymbol{m}'} = 0$ as well as $-\frac12 (\alpha^{-1} \beta)_{\boldsymbol{m} \boldsymbol{m}'} \neq 0$ to second order.

Therefore, $\ket{0}_G^\Lambda$ cannot be written as a product state over the $\boldsymbol{x}$-space sample point decomposition.
Hence we have formally established that there is entanglement between local degrees of freedom in the ground state of the field in the non-relativistic regime.
Now we investigate whether the entanglement manifests itself in certain entanglement measures for some simple situations where one can work out explicit expressions.
These will also serve to justify the conclusion of the above formal manipulations.
For the following computations, we will make use of the Gaussian state formalism, the relevant aspects of which we summarize in Appendix~\ref{apdx:gaussian}.

\subsection{Temperature of a single oscillator}\label{subsec:osc_temp}

First, we examine the reduced state of a single smeared field observable, with smearing function $f$.
We choose $f$ to be a general normalized ($\|f\|_2 = 1$) bandlimited smearing function.
Thus the subsystem is a single local oscillator generated by $\hat{\Phi}[f] := \int d\boldsymbol{x} f(\boldsymbol{x}) \hat{\Phi}(\boldsymbol{x})$ and $\hat{\Pi}[f] := \int d\boldsymbol{x} f(\boldsymbol{x}) \hat{\Pi}(\boldsymbol{x})$.

Here we will enforce the bandlimit via the smearing function.
For this and the following subsection, one could use either a spherically-symmetric cutoff, $\| \boldsymbol{k} \|_2 < \Lambda$, or a coordinate cutoff, $\| \boldsymbol{k} \|_\infty < \Lambda$.
The choice will only affect numerical prefactors.
Even though we made use of the latter cutoff in Subsection~\ref{subsec:sampling} to establish the form of the local degrees of freedom of the field in terms of sample points, if one restricts attention to finitely many sample points then the entanglement calculations can be performed in either case.
For a discussion of a similar situation, see \cite{Pye-Donnelly-Kempf2015}.

The reduced state on the subsystem defined by $f$ has a single symplectic eigenvalue,
\begin{equation}
  \nu = \frac{1}{\hbar} \Delta \Phi_f \Delta \Pi_f,
\end{equation}
where
\begin{align}
  \Delta \Phi_f^2 &:= \langle \hat{\Phi}[f]^2 \rangle = \int \frac{d\boldsymbol{k}}{(2\pi)^n} \frac{\hbar c^2}{2\omega_{\boldsymbol{k}}} | \tilde{f}(\boldsymbol{k}) |^2, \\
  \Delta \Pi_f^2 &:= \langle \hat{\Pi}[f]^2 \rangle = \int \frac{d\boldsymbol{k}}{(2\pi)^n} \frac{\hbar \omega_{\boldsymbol{k}}}{2c^2} | \tilde{f}(\boldsymbol{k}) |^2,
\end{align}
and where $\tilde{f}(\boldsymbol{k})$ is the Fourier transform of $f(\boldsymbol{x})$.
If $f$ has a bandlimit $\Lambda$, then it is straightforward to show that
\begin{equation}
  \nu = \frac12 + C (\Lambda/k_c)^4 + \mathcal{O}( (\Lambda/k_c)^6 ),
\end{equation}
where $C := \tfrac{1}{16} ( \tilde{f}_4 - \tilde{f}_2^2 )$ is a constant, and where
\begin{equation}
  \tilde{f}_p := \int \frac{d\boldsymbol{k}}{(2\pi)^n} ( \boldsymbol{k} / \Lambda )^p | \tilde{f}(\boldsymbol{k}) |^2.
\end{equation}
Note that generically we can choose $f$ such that $\tilde{f}_4 \neq \tilde{f}_2^2$, and hence $C \neq 0$.
For example, in the case of $n=1$, one can choose $f(x) = \sqrt{\frac{\Lambda}{\pi}} \sinc [ \Lambda (x-x_0) ]$ (i.e., the smearing function for a sample point centered at $x_0$), in which case $C = 1/180$.
One can then work out that the von Neumann entropy (corresponding to the entanglement entropy) of the reduced state on this single oscillator is, to lowest order,
\begin{equation}
  S \sim C \left( 1 - \log [ C (\Lambda/k_c)^4 ] \right) (\Lambda/k_c)^4.
\end{equation}
Note that this decays towards zero faster than $(\Lambda/k_c)^2$ as $\Lambda/k_c \to 0$, hence we consider this to be in the relativistic regime (the non-relativistic regime being contributions decaying slower or as slow as quadratically when $\Lambda/k_c \to 0$).

What can we conclude about the entanglement of a single oscillator?
On one hand, we showed that in the non-relativistic regime the ground state of the field is not fully separable over local degrees of freedom, yet here we see that nevertheless the entanglement entropy between one sample point and the rest of the system is zero in this regime.
However, this is simply an artifact of choosing the entropy to quantify entanglement.
In the following subsection we will calculate the logarithmic negativity between two sample points and show that it is non-zero, hence we can conclude that there is still entanglement in this regime.

However, for a single sample point we can still conclude that the state retains a non-zero temperature.
In analogy with the Unruh temperature characteristic of the Unruh effect, we can calculate a temperature for our single sample point using the symplectic eigenvalue we have calculated and identifying the coefficients from the thermal state decomposition in Eq.~\eqref{eq:gaussian_thermal_decomp} of Appendix~\ref{apdx:gaussian}:
\begin{equation}
  k_B T = \frac{\hbar \omega_f}{\log \left( \frac{\nu+1/2}{\nu-1/2} \right)}.
\end{equation}
Because the symplectic eigenvalue only determines the Boltzmann factor $e^{-\hbar \omega_f/k_B T}$, one has to choose an appropriate frequency scale, $\omega_f$, to distill a temperature from the Boltzmann factor.
An obvious choice would be to associate the energy scale $\hbar \omega_f$ with that of the Hamiltonian restricted to the oscillator defined by $f$, namely, $H_f = (c^2/2) \Pi[f]^2 + (\omega[f]^2/2c^2) \Phi[f]^2$ where $\omega[f]^2/c^2 := \| \boldsymbol{\nabla} f \|_2^2 + k_c^2$.
However, it is not possible to identify a suitable temperature so that the reduced density matrix of this oscillator takes the form $\frac1Z e^{-\beta \hat{H}_f}$.
Nevertheless, it is possible to define an effective frequency, $\omega_f$, so that the reduced density matrix is a thermal state of an effective Hamiltonian of the form $(c^2/2) \Pi[f]^2 + (\omega_f^2/2c^2) \Phi[f]^2$.
One can identify this frequency by writing the ground state of the full system and directly computing the reduced state onto the subsystem defined by $f$.
This is a straightforward calculation and we will omit it here (for a reference, see \cite{Bombelli-Koul-Lee-Sorkin1986}).
We find that one should choose $\omega_f = c^2 \Delta \Pi_f / \Delta \Phi_f$, although we shall see that any other choice will suffice, provided it agrees with the Compton frequency $\omega_c = mc^2/\hbar$ to lowest order in the expansion in $\Lambda/k_c$.

The above expression for $k_B T$ is not analytic at $\Lambda/k_c = 0$, nevertheless one can show that the expression decays slower than any polynomial as $\Lambda/k_c \to 0$, i.e., $\lim_{\Lambda/k_c \to 0} (\Lambda/k_c)^n/(k_B T) = 0$ for $n>0$.
Therefore, we can conclude that the temperature of a single local oscillator is non-zero in the non-relativistic regime.
In particular, to leading order,
\begin{equation}
  k_B T \sim \frac{mc^2}{ \log [ \left( \Lambda/k_c \right)^{-4} ] },
\end{equation}
where, in standard fashion, we use $f(x) \sim g(x)$ to indicate $\lim_{x \to 0} f(x)/g(x) = 1$.
Note that this leading order term does not depend on the particular choice of the smearing function $f$ (apart from the condition that it is bandlimited), nor the spatial dimension, nor the particular choice of frequency $\omega_f$ (provided it is analytic and equals $\omega_c$ at zeroth order).

\subsection{Logarithmic negativity between two local oscillators}\label{subsec:osc_logneg}

To demonstrate that there is a non-trivial measure of entanglement in the non-relativistic regime of the field theory, here we will examine the logarithmic negativity between two local degrees of freedom of the field.
These local degrees of freedom will be defined by two smearing functions, $f_1$ and $f_2$, which we will assume are bandlimited and are orthonormal, so that
\begin{equation}
  ( f_i | f_j ) = \int d\boldsymbol{x} f_i(\boldsymbol{x}) f_j(\boldsymbol{x}) = \int \frac{d\boldsymbol{k}}{(2\pi)^n} \tilde{f}^\ast_i(\boldsymbol{k}) \tilde{f}_j(\boldsymbol{k}) = \delta_{ij}.
\end{equation}
For simplicity, we will also assume that $|\tilde{f}_1(\boldsymbol{k})| = |\tilde{f}_2(\boldsymbol{k})|$, so that the local uncertainty of the oscillators are the same, i.e., $\Delta \Phi_f^2 \equiv \Delta \Phi_{f_1}^2 = \Delta \Phi_{f_2}^2$.
Similarly, we will have $\Delta \Pi_f^2 \equiv \Delta \Pi_{f_1}^2 = \Delta \Pi_{f_2}^2$.
This assumption is made only to simplify the form of the result, and will suffice for the purposes of the current demonstration.
Intuitively this requirement is stating that the two subsystems are simply translated versions of one another; for example, in the bandlimited theory these could be $\tilde{f}_i(\boldsymbol{k}) = \left( \frac{\pi}{\Lambda} \right)^{n/2} e^{-i \boldsymbol{k} \cdot \boldsymbol{x}_i} \chi( \| \boldsymbol{k} \|_\infty < \Lambda)$, where $\boldsymbol{x}_1 - \boldsymbol{x}_2 = \boldsymbol{N}\pi/\Lambda$ and $\boldsymbol{N} \in \mathds{Z}^n$ (i.e., $\boldsymbol{x}_1$ is $\boldsymbol{x}_2$ translated by an integral number of lattice spacings).

The correlation matrix for the Klein-Gordon field theory, restricted to these two oscillators, is:
\begin{equation}
  \Sigma =
  \begin{bmatrix}
    \Delta \Phi_f^2 & 0 & \Phi_{12} & 0 \\
    0 & \Delta \Pi_f^2 & 0 & \Pi_{12} \\
    \Phi_{12} & 0 & \Delta \Phi_f^2 & 0 \\
    0 & \Pi_{12} & 0 & \Delta \Pi_f^2
  \end{bmatrix},
\end{equation}
where
\begin{align}
  \Delta \Phi_f^2 &= \int \frac{d\boldsymbol{k}}{(2\pi)^n} \frac{\hbar c^2}{2\omega_{\boldsymbol{k}}} | \tilde{f}(\boldsymbol{k}) |^2, \\
  \Phi_{12} &:= \int \frac{d\boldsymbol{k}}{(2\pi)^n} \frac{\hbar c^2}{2\omega_{\boldsymbol{k}}} \text{Re}[ \tilde{f}^\ast_1(\boldsymbol{k}) \tilde{f}_2(\boldsymbol{k})  ], \\
  \Delta \Pi_f^2 &= \int \frac{d\boldsymbol{k}}{(2\pi)^n} \frac{\hbar \omega_{\boldsymbol{k}}}{2c^2} | \tilde{f}(\boldsymbol{k}) |^2, \\
  \Pi_{12} &:= \int \frac{d\boldsymbol{k}}{(2\pi)^n} \frac{\hbar \omega_{\boldsymbol{k}}}{2c^2} \text{Re}[ \tilde{f}^\ast_1(\boldsymbol{k}) \tilde{f}_2(\boldsymbol{k})  ],
\end{align}
and we have denoted $|\tilde{f}(\boldsymbol{k})| \equiv |\tilde{f}_1(\boldsymbol{k})| = |\tilde{f}_2(\boldsymbol{k})|$.

In order to calculate the logarithmic negativity, we need to determine the symplectic eigenvalues of the partially-transposed covariance matrix, which in this case amounts to replacing $\Pi_{12} \mapsto -\Pi_{12}$ in the above matrix.
We find that the two symplectic eigenvalues are
\begin{equation}
  \nu_\pm = \frac{1}{\hbar} \sqrt{(\Delta \Phi_f^2 \pm \Phi_{12})(\Delta \Pi_f^2 \pm \Pi_{12})}.
\end{equation}
If we expand this expression in powers of $\Lambda / k_c$, we find,
\begin{equation}
  \nu_\pm = \frac12 \mp \frac14 \tilde{F}_{12}^{(2)} (\Lambda/k_c)^2 + \mathcal{O} ( (\Lambda/k_c)^4 ),
\end{equation}
where
\begin{equation}
  \tilde{F}_{12}^{(2)} := \int \frac{d\boldsymbol{k}}{(2\pi)^n} (\boldsymbol{k}/\Lambda)^2 \hspace{1mm} \text{Re}[ \tilde{f}^\ast_1(\boldsymbol{k}) \tilde{f}_2(\boldsymbol{k})  ].
\end{equation}
Hence, the logarithmic negativity between these two subsystems is nonzero to second order in $\Lambda/k_c$,
\begin{equation}
  E_N \sim \frac12 | \tilde{F}_{12}^{(2)} | (\Lambda/k_c)^2.
\end{equation}
Therefore, we can conclude also from this calculation that these subsystems are entangled.
For example, in $n=1$ dimensions, we can choose smearings $\tilde{f}_i(k) = \sqrt{\frac{\pi}{\Lambda}} e^{-i k x_i} \chi(|k|<\Lambda)$, which give $E_N \sim \frac{1}{N^2 \pi^2} (\Lambda/k_c)^2$, where $N$ is the number of lattice spacings between sample points $x_1$ and $x_2$.

\section{Conclusion and Outlook}

The main goal of this paper was to elucidate the obstructions to localizability of particle states in quantum field theory by attempting to recover the known localizability properties of wavefunctions in non-relativistic quantum mechanics under a non-relativistic approximation.
We enacted the non-relativistic approximation by identifying a subspace of the global Hilbert space corresponding to states satifying an ultraviolet cutoff set by the Compton scale.
We showed that one can recover many of the characteristic features of NRQM beyond the Schr\"odinger equation.
However, in this study we have also identified remaining localizability issues within this imitation of NRQM, such as differences between the standard and non-relativistic localization schemes in this regime, the interpretation of measurements associated with localizing the wavefunction, and non-locality due to the bandlimit and sampling properties.

Finally, we showed that the lingering discrepancy between the two localization schemes leads to the survival of ground state entanglement in the non-relativistic approximation \updated{of a Klein-Gordon field}.
Hence, we can conclude that the existence of the Unruh effect does not rely on the presence of relativistic effects \updated{(insofar as it is related to ground state entanglement)}.
Indeed, despite the observation that the Unruh temperature vanishes in the naive limit $c \to \infty$, we demonstrated that the local degrees of freedom exhibit a non-zero temperature in the non-relativistic approximation.
However, the fact that ground state entanglement \updated{of a Klein-Gordon field} is not relativistic could support the use of non-relativistic detector systems to probe this entanglement (see, e.g., \cite{Valentini1991,Reznik2003,PozasKerstjens-MartinMartinez2015}).
That is, the non-relativistic detector systems would not have to enter the relativistic regime in order to access the entanglement.
This may also suggests that the alternative derivation of the Unruh effect using accelerating detectors would survive the non-relativistic approximation.

The remaining localizability issues we have identified warrant further investigation.
For example, it would be interesting to further explore the measurement theory for the operators used to characterize the localizability of the wavefunctions in the non-relativistic approximation.
One task is to find the Kraus operators associated with the POVM, in order to determine the appropriate state-update rule.
Perhaps this can be achieved through a dilation, for example, using the modes above the cutoff.
As we mentioned, these operators also arise in simple quantum-mechanical systems, hence it seems such a clarification would be broadly applicable.

Also, the intrinsic slowly-decaying non-locality of the wavefunctions indicates that the requirement of an ultraviolet cutoff for the non-relativistic approximation may have been too severe.
This requirement was predicated on the assumption that one should recover a Hilbert space for the description of NRQM.
One could consider attempting to relax this requirement, for example, by choosing a space consisting of the (finite) span of a collection of states which decay sufficiently quickly in momentum space (e.g., Gaussian smearing functions).
The issue with such an approach is that one would lose much of the structure of NRQM (and helpful mathematical tools) without the Hilbert space assumption.
For instance, one would retain the superposition principle, but lose much of the functional analytic structure, such as the spectral theorem.
Perhaps this is an appropriate compromise in order to resolve certain localizability issues.
However, to obtain a model with a fixed number of particles, one should ensure the prevention of particle creation, despite allowing modes above the cutoff.

We note also that the classical limit of the Klein-Gordon quantum field theory is formally similar to the non-relativistic approximation, as one is also considering a regime where the Compton wavenumber, $k_c = mc/\hbar$, is large compared to some other scale.
This raises the question of whether the classical regime of the Klein-Gordon theory should also be considered a bandlimited theory.

Lastly, given that the local degrees of freedom of the field theory remain entangled in the non-relativistic regime, it is natural to ask about the fate of the Reeh-Schlieder theorem.
The Reeh-Schlieder theorem establishes the cyclicity of the ground state for the local operators confined to any open region of spacetime.
Although this can be linked to ground state entanglement, the presence of entanglement is not sufficient to demonstrate this fact.
Determining whether there is an analogue of the Reeh-Schlieder theorem in this regime may help further elucidate the persisting localizability issues in the non-relativistic regime.

\section*{Acknowledgements}

The authors would like to thank their supervisor Achim Kempf for his support and the very helpful discussions.
MP would like to thank Eduardo Mart\'in-Mart\'inez also for his support and the stimulating discussions.
MP would also like to thank Charis Anastopoulos, Doreen Fraser, and Juan Le\'on for sharing valuable insight.
The authors also thank Jos\'e de Ram\'on for the constructive feedback and the psychological support.
JP acknowledges support from the Natural Sciences and Engineering Research Council (NSERC) of Canada through the Doctoral Canada Graduate Scholarship (CGS) program, as well as from the Ontario Graduate Scholarship (OGS) program.
This research was supported in part by Perimeter Institute for Theoretical Physics. Research at Perimeter Institute is supported by the Government of Canada through the Department of Innovation, Science and Economic Development and by the Province of Ontario through the Ministry of Research and Innovation.

\appendix

\section{Bogoliubov transformations and unitary inequivalence}\label{apdx:bbv}

\subsection{Bogoliubov transformations}

A general Bogoliubov transformation between two sets of annihilation and creation operators takes the form:
\begin{equation}\label{eq:genl_bbv}
  \hat{a}_i = \sum_j ( \alpha_{ij} \hat{b}_j + \beta_{ij} \hat{b}_j^\dagger ).
\end{equation}
So that both sets of annihilation and creation operators satisfy the canonical commutation relations, the operators $\alpha$ and $\beta$ must satisfy the conditions,
\begin{equation}
  \alpha \alpha^\dagger - \beta \beta^\dagger = \mathds{1}, \quad \text{and} \quad \alpha \beta^T = \beta \alpha^T.
\end{equation}
Then the inverse transformation is given by
\begin{equation}
  \hat{b}_j = \sum_i ( \alpha_{ij}^\ast \hat{a}_i - \beta_{ij} \hat{a}_i^\dagger ).
\end{equation}

For the case of the local and global operators introduced in Section~\ref{sec:background}, we have
\begin{align}
  \hat{a}_{\boldsymbol{k}} &:= \sqrt{\frac{\omega_{\boldsymbol{k}}}{2\hbar c^2}} \hat{\Phi}_{\boldsymbol{k}} + i \sqrt{\frac{c^2}{2\hbar \omega_{\boldsymbol{k}}}} \hat{\Pi}_{\boldsymbol{k}} \nonumber \\
  &= \sqrt{\frac{\omega_{\boldsymbol{k}}}{2\hbar c^2}} \int d\boldsymbol{x} \hspace{1mm} e^{-i \boldsymbol{k} \cdot \boldsymbol{x}} \sqrt{\frac{\hbar^2}{2m}} ( \hat{b}_{\boldsymbol{x}} + \hat{b}_{\boldsymbol{x}}^\dagger ) + i \sqrt{\frac{c^2}{2\hbar \omega_{\boldsymbol{k}}}} \int d\boldsymbol{x} \hspace{1mm} e^{-i \boldsymbol{k} \cdot \boldsymbol{x}} (-i) \sqrt{\frac{m}{2}} ( \hat{b}_{\boldsymbol{x}} - \hat{b}_{\boldsymbol{x}}^\dagger ) \nonumber \\
  &= \int d\boldsymbol{x} \hspace{1mm} e^{-i \boldsymbol{k} \cdot \boldsymbol{x}} \left[ \frac12 \left( \sqrt{\frac{\hbar \omega_{\boldsymbol{k}}}{mc^2}} + \sqrt{\frac{mc^2}{\hbar \omega_{\boldsymbol{k}}}} \right) \hat{b}_{\boldsymbol{x}} + \frac12 \left( \sqrt{\frac{\hbar \omega_{\boldsymbol{k}}}{mc^2}} - \sqrt{\frac{mc^2}{\hbar \omega_{\boldsymbol{k}}}} \right) \hat{b}_{\boldsymbol{x}}^\dagger \right].
\end{align}
Hence, the Bogoliubov coefficients for this transformation are
\begin{equation}
  \alpha_{\boldsymbol{k} \boldsymbol{x}} := e^{-i \boldsymbol{k} \cdot \boldsymbol{x}} c_+(\boldsymbol{k}), \quad \text{and} \quad \beta_{\boldsymbol{k} \boldsymbol{x}} := e^{-i \boldsymbol{k} \cdot \boldsymbol{x}} c_-(\boldsymbol{k}),
\end{equation}
where
\begin{equation}
  c_\pm(\boldsymbol{k}) := \frac12 \left( \sqrt{ \hbar \omega_{\boldsymbol{k}} / mc^2 } \pm \sqrt{ mc^2 / \hbar \omega_{\boldsymbol{k}}} \right)
\end{equation}
We also have the inverse transformation,
\begin{equation}
  \hat{b}_{\boldsymbol{x}} = \int \frac{d\boldsymbol{k}}{(2\pi)^n} \left[ e^{i \boldsymbol{k} \cdot \boldsymbol{x}} c_+(\boldsymbol{k}) \hat{a}_{\boldsymbol{k}} - e^{-i \boldsymbol{k} \cdot \boldsymbol{x}} c_-(\boldsymbol{k}) \hat{a}_{\boldsymbol{k}}^\dagger \right].
\end{equation}
For the purposes of this paper, we are considering the relativistic dispersion relation $\omega_{\boldsymbol{k}} = c \sqrt{ \boldsymbol{k} + k_c^2 }$, where $k_c := mc/\hbar$, in which case $c_\pm(\boldsymbol{k}) = \tfrac12 [ ( 1 + (\boldsymbol{k}/k_c)^2 )^{1/4} \pm ( 1 + (\boldsymbol{k}/k_c)^2 )^{-1/4} ]$.
The form of the above Bogoliubov transformation would also hold for more general dispersion relations, e.g., in cases where one includes higher powers of the spatial derivatives.
This situation arises, for example, with condensed matter systems, sonic analogues \cite{Unruh1981,Unruh1995}, and effective Planck-scale corrections \cite{AmelinoCameliaEtal1998,Gambini-Pullin1999,Magueijo-Smolin2002,Myers-Pospelov2003,GirelliEtal2007}.
We see that regardless of the dispersion relation used, provided it is not trivially $\hbar \omega_{\boldsymbol{k}} = mc^2$, the above $\beta$-coefficients will not vanish.
Thus, modified dispersion relations generically will have qualitatively minor differences in, e.g., entanglement considerations as those examined in Section~\ref{sec:entanglement}.
There we show there will be non-trivial entanglement between spatial degrees of freedom provided that the operator $(\alpha^{-1} \beta)$ has non-zero off-diagonal components.
Changing the dispersion relation will generally only have the effect of changing numerical prefactors.
This is consistent with previous investigations into the Unruh effect and ground state entanglement with modified dispersion relations (see, e.g., \cite{Unruh1981,Unruh1995,Corley-Jacobson1996,Corley1998,Nesterov-Solodukhin2011,Solodukhin2011}).

Similar computations can be done for the bandlimited operators introduced in Section~\ref{sec:NR_limit}.
Here we will use the cutoff $\| \boldsymbol{k} \|_\infty < \Lambda$ described in Subsection~\ref{subsec:sampling} in order to discuss the operators at the sampling points, $\hat{b}_{\boldsymbol{m}} \equiv \left( \frac{\pi}{\Lambda} \right)^{n/2} \hat{b}_{\boldsymbol{x}_{\boldsymbol{m}}^{(\boldsymbol{\alpha})}}$.
The Bogoliubov transformation between these and the global operators is,
\begin{equation}
  \hat{a}_{\boldsymbol{k}} = \left( \frac{\pi}{\Lambda} \right)^{n/2} \sum_{\boldsymbol{m} \in \mathds{Z}^n} e^{-i \boldsymbol{k} \cdot \boldsymbol{x}_{\boldsymbol{m}}^{(\boldsymbol{\alpha})}} \left[ \hat{b}_{\boldsymbol{m}} + \frac14 (\boldsymbol{k}/k_c)^2 \hat{b}_{\boldsymbol{m}}^\dagger \right],
\end{equation}
with inverse,
\begin{equation}
  \hat{b}_{\boldsymbol{m}} = \left( \frac{\pi}{\Lambda} \right)^{n/2} \int_{\| \boldsymbol{k} \|_\infty < \Lambda} \frac{d\boldsymbol{k}}{(2\pi)^n} \left[ e^{i \boldsymbol{k} \cdot \boldsymbol{x}_{\boldsymbol{m}}^{(\boldsymbol{\alpha})}} \hat{a}_{\boldsymbol{k}} - e^{-i \boldsymbol{k} \cdot \boldsymbol{x}_{\boldsymbol{m}}^{(\boldsymbol{\alpha})}} \frac14 (\boldsymbol{k}/k_c)^2 \hat{a}_{\boldsymbol{k}}^\dagger \right].
\end{equation}
The Bogoliubov coefficients after the bandlimitation are:
\begin{equation}
  \alpha^\Lambda_{\boldsymbol{k} \boldsymbol{m}} := \left( \frac{\pi}{\Lambda} \right)^{n/2} e^{-i \boldsymbol{k} \cdot \boldsymbol{x}_{\boldsymbol{m}}^{(\boldsymbol{\alpha})}}, \quad \text{and} \quad \beta^\Lambda_{\boldsymbol{k} \boldsymbol{m}} := \left( \frac{\pi}{\Lambda} \right)^{n/2} e^{-i \boldsymbol{k} \cdot \boldsymbol{x}_{\boldsymbol{m}}^{(\boldsymbol{\alpha})}} \frac14 (\boldsymbol{k}/k_c)^2.
\end{equation}

\subsection{Unitary inequivalence}

Bogoliubov transformations describe changes between different sets of annihilation and creation operators.
These sets of operators can each be used to generate a Fock space, as described in Section~\ref{sec:background}.
It is natural to ask whether these Fock spaces are unitarily equivalent representations of the algebra generated by these operators.

Let us consider a general Bogoliubov transformation of the form \eqref{eq:genl_bbv}.
Let us denote the $a$-Fock space as $\mathcal{F}_a$, with vacuum $\ket{0}_a$ and total number operator $\hat{N}_a := \sum_i \hat{a}_i^\dagger \hat{a}_i$.
Similar for the $b$-Fock space.
These Fock spaces are unitarily equivalent representations if and only if
\begin{equation}
  \left._a \hspace{-1mm} \bra{0} \hat{N}_b \ket{0}_a \right. = \left._b \hspace{-1mm} \bra{0} \hat{N}_a \ket{0}_b \right. = \sum_{ij} | \beta_{ij} |^2 < \infty,
\end{equation}
i.e., if the $a$-vacuum contains finitely many $b$-particles, and vice-versa \cite{Fulling1989,Wald1994}.
It can also be stated as the requirement that the coefficients, $\beta_{ij}$, define the components of a Hilbert-Schmidt operator.
Imposing the Bogoliubov transformation as an abstract relationship between the two sets of operators dictates that in a representation the two vacua must be related by
\begin{equation}
  \ket{0}_a = N e^{-\frac12 \sum_{j,k} (\alpha^{-1} \beta)_{jk} \hat{b}_j^\dagger \hat{b}_k^\dagger} \ket{0}_b,
\end{equation}
The above condition for unitary equivalence is the same as the condition that the expression on the right-hand side of the above equation is normalizable, hence a valid state which can be identified with $\ket{0}_a$ \cite{Fulling1989}.

Applying this to our cases, we can easily see that the local and global Fock spaces, $\mathcal{F}_L$ and $\mathcal{F}_G$ from Section~\ref{sec:background}, are unitarily inequivalent, since
\begin{equation}
  \int d\boldsymbol{x} \int \frac{d\boldsymbol{k}}{(2\pi)^n} | \beta_{\boldsymbol{k} \boldsymbol{x}} |^2 = \int d\boldsymbol{x} \int \frac{d\boldsymbol{k}}{(2\pi)^n} | c_-(\boldsymbol{k}) |^2 = \infty,
\end{equation}
due to the divergence of the $\boldsymbol{k}$-integration as well as the infinite volume of space.
After introducing the cutoff $\| \boldsymbol{k} \|_\infty < \Lambda$, the unitary inequivalence between the (bandlimited) local and global operators does not subside.
Indeed, we see
\begin{equation}
  \sum_{\boldsymbol{m}} \int_{\| \boldsymbol{k} \|_\infty < \Lambda} \frac{d\boldsymbol{k}}{(2\pi)^n} | \beta^\Lambda_{\boldsymbol{k} \boldsymbol{m}} |^2 = \sum_{\boldsymbol{m}} \int_{\| \boldsymbol{k} \|_\infty < \Lambda} \frac{d\boldsymbol{k}}{(2\pi)^n} | c_-(\boldsymbol{k}) |^2 = \infty.
\end{equation}
In this case, we see that the divergence is solely due to the infinite volume of space.
This could clearly be tamed by considering a field theory in a finite volume of space, since in this case with an ultraviolet and infrared cutoff there would only be finitely many degrees of freedom, so the Stone-von Neumann theorem would apply.

\section{Gaussian state formalism}\label{apdx:gaussian}

For the calculations undertaken in Section~\ref{sec:entanglement}, we exploit the fact that the ground state of a free Klein-Gordon field is a continuous variable Gaussian state, hence we can rely on the well-developed Gaussian-state formalism for continuous variable quantum information.
For a general reference on continuous variable Gaussian methods, see, for example, \cite{Serafini2017}.
We will summarize the necessary tools here.

We will generally denote phase space vectors as $\boldsymbol{\hat{r}} := ( \hat{\Phi}_1, \hat{\Pi}_2, \hat{\Phi}_2, \hat{\Pi}_2, \dots )^T$, where $\hat{\Phi}_i := \hat{\Phi}[f_i] = \int d\boldsymbol{x} f_i(\boldsymbol{x}) \hat{\Phi}(\boldsymbol{x})$ and $\hat{\Pi}_i := \hat{\Pi}[f_i] = \int d\boldsymbol{x} f_i(\boldsymbol{x}) \hat{\Pi}(\boldsymbol{x})$.
For the states we will consider, it will always be the case that $\langle \hat{r}_i \rangle = 0$.
We will also denote the covariance matrix by $\Sigma_{ij} := \tfrac12 \langle \hat{r}_i \hat{r}_j + \hat{r}_j \hat{r}_i \rangle$, and the symplectic matrix by $i \Omega_{ij} \mathds{1} := [ \hat{r}_i, \hat{r}_j ]$.
These two matrices completely characterize the density matrix of a Gaussian state (along with $\langle \boldsymbol{\hat{r}} \rangle$ if it is non-zero).

Reduced states of subsystems of a Gaussian states are always Gaussian, and are characterized by the projections of these matrices onto the corresponding subspace of the phase space.
For example, for the subsystem corresponding to finitely many smearings, $f_1, \dots, f_N$, the projectors are simply those onto the $2N$-dimensional subspace of phase space vectors of the form $( \hat{\Phi}_1, \hat{\Pi}_1, \dots, \hat{\Phi}_N, \hat{\Pi}_N )$.
If we denote the projectors onto this subsystem by $P_S$, then the covariance and symplectic matrices characterizing the state of this subsystem are: $\Sigma_S := P_S \Sigma P_S$ and $\Omega_S := P_S \Omega P_S$.
These are convenient for calculations because they are finite-dimensional matrices which completely determine the reduced state.
In particular, we do not have to appeal to the form of the complementary components of the matrices for the state of the full system.

For Gaussian states, many quantities of interest are readily expressible in terms of the symplectic eigenvalues of the covariance matrix, i.e., eigenvalues of the matrix $\Omega^{-1} \Sigma$ (or often the eigenvalues of $\Omega_S^{-1} \Sigma_S$ corresponding to a reduced state).
These eigenvalues always take the form $\{ \pm i \nu_j \}_j$, with $\nu_j \geq 1/2$.
The usefulness in calculating these symplectic eigenvalues is due to the fact that one can show on general grounds that a Gaussian density matrix can always be brought to a form where it can be expressed as a product of thermal density matrices, which are each parametrized by one of the symplectic eigenvalues:
\begin{equation}\label{eq:gaussian_thermal_decomp}
  \rho = \otimes_j \rho_j, \qquad \rho_j := \frac{1}{\nu_j+1/2} \sum_{n=0}^\infty \left( \frac{\nu_j-1/2}{\nu_j+1/2} \right)^n \ket{n} \bra{n}.
\end{equation}
Then, for example, one can calculate the von Neumann entropy of the state as
\begin{equation}
  S(\rho) = - \tr ( \rho \log \rho ) = \sum_j [ (\nu_j+1/2) \log (\nu_j+1/2) - (\nu_j-1/2) \log (\nu_j-1/2) ].
\end{equation}

Another quantity of interest for us is the logarithmic negativity of a bipartite Gaussian state.
This is defined as $E_N(\rho) := \log \| \tilde{\rho} \|_1$, where $\| \cdot \|_1$ denotes the 1-norm (sum of the absolute values of the eigenvalues) and $\tilde{\rho}$ denotes the partial transpose of $\rho$ (obtained by applying the matrix transpose to one of the two subsystems in the bipartition).
In phase space, the partial transpose is represented as a reflection of the momenta of one of the two subsystems.
For example, for two modes, the reflection acts on the phase space vectors as: $( \hat{\Phi}_1, \hat{\Pi}_1, \hat{\Phi}_2, \hat{\Pi}_2 ) \mapsto ( \hat{\Phi}_1, \hat{\Pi}_1, \hat{\Phi}_2, -\hat{\Pi}_2 )$.
Therefore, this operation is simple to apply to the matrices which define the Gaussian state.
To calculate the logarithmic negativity, one then obtains the symplectic eigenvalues of the covariance matrix after applying this reflection, from which it is easy to show that
\begin{equation}
  E_N(\rho) = \sum_j \max \{ 0, -\log(2\tilde{\nu}_j) \},
\end{equation}
where we denote by $\{ \tilde{\nu}_j \}_j$ the symplectic eigenvalues of the partially-transposed covariance matrix.
Note that since the partially-transposed density matrix is not necessarily a valid density matrix, one does not generally have that $\tilde{\nu}_j \geq 1/2$ (otherwise the logarithmic negativity would always vanish), although we still have $\tilde{\nu}_j > 0$.

Generally there are (bound) entangled states for which the logarithmic negativity is zero, although a non-zero $E_N$ always implies that the state is entangled across the bipartition.
Furthermore, in the case where one of the two subsystems consists of a single mode, $E_N$ is non-zero if and only if the two subsystems are entangled \cite{Serafini2017,Simon2000,Werner-Wolf2001}.
In Section~\ref{sec:entanglement} we are considering such a case.

\bibliographystyle{ieeetr}
\bibliography{main}

\end{document}